\documentclass[twocolumn]{aastex62}
\usepackage{hyperref}
\usepackage[T1]{fontenc}
\usepackage{apjfonts} 

\usepackage[hang,flushmargin]{footmisc} 

\usepackage{url}

\hypersetup{
    allcolors = {blue},
}

\newcommand{\DSFP}{\altaffiliation{LSSTC Data Science Fellow}}
\newcommand{\CfA}{\affiliation{Center for Astrophysics \textbar{} Harvard \& Smithsonian, 60 Garden Street, Cambridge, MA 02138-1516, USA}}
\newcommand{\Hubble}{\altaffiliation{Hubble Fellow}}
\newcommand{\Carnegie}{\affiliation{Observatories of the Carnegie Institute for Science, 813 Santa Barbara Street, Pasadena, CA 91101-1232, USA}}
\newcommand{\Edinburgh}{\affiliation{Institute for Astronomy, University of Edinburgh, Royal Observatory, Blackford Hill, EH9 3HJ, UK}}
\newcommand{\Birmingham}{\affiliation{Birmingham Institute for Gravitational Wave Astronomy and School of Physics and Astronomy, University of Birmingham, Birmingham B15 2TT, UK}}
\newcommand{\CIERA}{\affiliation{Center for Interdisciplinary Exploration and Research in Astrophysics and Department of Physics and Astronomy, \\Northwestern University, 2145 Sheridan Road, Evanston, IL 60208-3112, USA}}
\newcommand{\Ohio}{\affiliation{Astrophysical Institute, Department of Physics and Astronomy, 251B Clippinger Lab, Ohio University, Athens, OH 45701-2942, USA}}
\newcommand{\Geneva}{\affiliation{ISDC, Department of Astronomy, University of Geneva, Chemin d'\'Ecogia, 16 CH-1290 Versoix, Switzerland}}
\newcommand{\Einstein}{\altaffiliation{Einstein Fellow}}
\newcommand{\INPE}{\affiliation{Instituto Nacional de Pesquisas Espaciais, Avenida dos Astronautas 1758, 12227-010, S\~ao Jos\'e dos Campos -- SP, Brazil}}
\newcommand{\CTIO}{\affiliation{Cerro Tololo Inter-American Observatory, National Optical Astronomy Observatory, Casilla 603, La Serena, Chile}}
\newcommand{\Bath}{\affiliation{Department of Physics, University of Bath, Claverton Down, Bath, BA2 7AY, UK}}
\newcommand{\MMT}{\affiliation{MMT and Steward Observatories, University of Arizona, 933 North Cherry Avenue, Tucson, AZ 85721-0065, USA}}
\newcommand{\Potsdam}{\affiliation{Institut f\"ur Physik und Astronomie, Universit\"at Potsdam, Haus 28, Karl-Liebknecht-Str. 24/25, D-14476 Potsdam-Golm, Germany}}
\newcommand{\UNC}{\affiliation{Department of Physics and Astronomy, University of North Carolina, 120 East Cameron Avenue, Chapel Hill, NC, 27599, USA}}
\newcommand{\AAS}{\affiliation{American Astronomical Society, 1667 K~Street NW, Suite 800, Washington, DC 20006-1681, USA}}

\shorttitle{MMT/SOAR Follow-up of GW Events S190425z and S190426c}
\shortauthors{Hosseinzadeh, Cowperthwaite, Gomez, Villar, Nicholl, Margutti, et al.}

\begin{document}

\title{Follow-up of the Neutron Star Bearing Gravitational-wave Candidate Events S190425z and S190426c with MMT and SOAR}

\correspondingauthor{Griffin Hosseinzadeh}
\email{griffin.hosseinzadeh@cfa.harvard.edu}

\author[0000-0002-0832-2974]{G.~Hosseinzadeh}
\DSFP\CfA

\author[0000-0002-2478-6939]{P.~S.~Cowperthwaite}
\Hubble\Carnegie

\author[0000-0001-6395-6702]{S.~Gomez}
\CfA

\author[0000-0002-5814-4061]{V.~A.~Villar}
\CfA

\author[0000-0002-2555-3192]{M.~Nicholl}
\Edinburgh\Birmingham

\author[0000-0003-4768-7586]{R.~Margutti}
\CIERA

\collaboration{({\it These authors contributed equally to this work.})}

\author[0000-0002-9392-9681]{E.~Berger}
\CfA
\nocollaboration

\author[0000-0002-7706-5668]{R.~Chornock}
\Ohio

\author[0000-0001-8340-3486]{K.~Paterson}
\CIERA

\author[0000-0002-7374-935X]{W.~Fong}
\CIERA

\author[0000-0001-6353-0808]{V.~Savchenko}
\Geneva

\author[0000-0002-5096-9464]{P.~Short}
\Edinburgh

\author[0000-0002-8297-2473]{K.~D.~Alexander}
\Einstein\CIERA

\author[0000-0003-0526-2248]{P.~K.~Blanchard}
\CfA

\author[0000-0003-3338-3186]{J.~Braga}
\INPE

\author[0000-0002-2830-5661]{M.~L.~Calkins}
\CfA

\author[0000-0003-4553-4033]{R. Cartier}
\CTIO

\author[0000-0001-5126-6237]{D.~L.~Coppejans}
\CIERA

\author[0000-0003-0307-9984]{T.~Eftekhari}
\CfA

\author[0000-0003-1792-2338]{T.~Laskar}
\Bath

\author[0000-0002-4245-2318]{C.~Ly}
\MMT

\author[0000-0002-7640-236X]{L.~Patton}
\CfA

\author[0000-0003-4615-6556]{I.~Pelisoli}
\Potsdam

\author[0000-0002-5060-3673]{D.~E.~Reichart}
\UNC

\author[0000-0003-0794-5982]{G.~Terreran}
\CIERA

\author[0000-0003-3734-3587]{P.~K.~G.~Williams}
\CfA\AAS

\received{2019 May 6}
\revised{2019 June 3}
\accepted{2019 June 3}
\published{2019 July 18}

\begin{abstract}
On 2019 April 25.346 and 26.640 UT the Laser Interferometer Gravitational-Wave Observatory and the Virgo gravitational-wave (GW) observatory announced the detection of the first candidate events in Observing Run 3 that contained at least one neutron star (NS). S190425z is a likely binary neutron star (BNS) merger at $d_L = 156 \pm 41$ Mpc, while S190426c is possibly the first NS--black hole (BH) merger ever detected, at $d_L = 377 \pm 100$ Mpc, although with marginal statistical significance. Here we report our optical follow-up observations for both events using the MMT 6.5~m telescope, as well as our spectroscopic follow-up of candidate counterparts (which turned out to be unrelated) with the 4.1~m SOAR telescope.  We compare to publicly reported searches, explore the overall areal coverage and depth, and evaluate those in relation to the optical/near-infrared (NIR) kilonova emission from the BNS merger GW170817, to theoretical kilonova models, and to short gamma-ray burst (SGRB) afterglows.  We find that for a GW170817-like kilonova, the partial volume covered spans up to about 40\% for S190425z and 60\% for S190426c. For an on-axis jet typical of SGRBs, the search effective volume is larger, but such a configuration is expected in at most a few percent of mergers.  We further find that wide-field $\gamma$-ray and X-ray limits rule out luminous on-axis SGRBs, for a large fraction of the localization regions, although these searches are not sufficiently deep in the context of the $\gamma$-ray emission from GW170817 or off-axis SGRB afterglows.  The results indicate that some optical follow-up searches are sufficiently deep for counterpart identification to about 300 Mpc, but that localizations better than 1000 deg$^2$ are likely essential.
\end{abstract}


\section{Introduction}
\label{sec:intro}

\begin{figure*}[t!]
\begin{center}
\scalebox{1.}
{\includegraphics[width=0.6\textwidth]{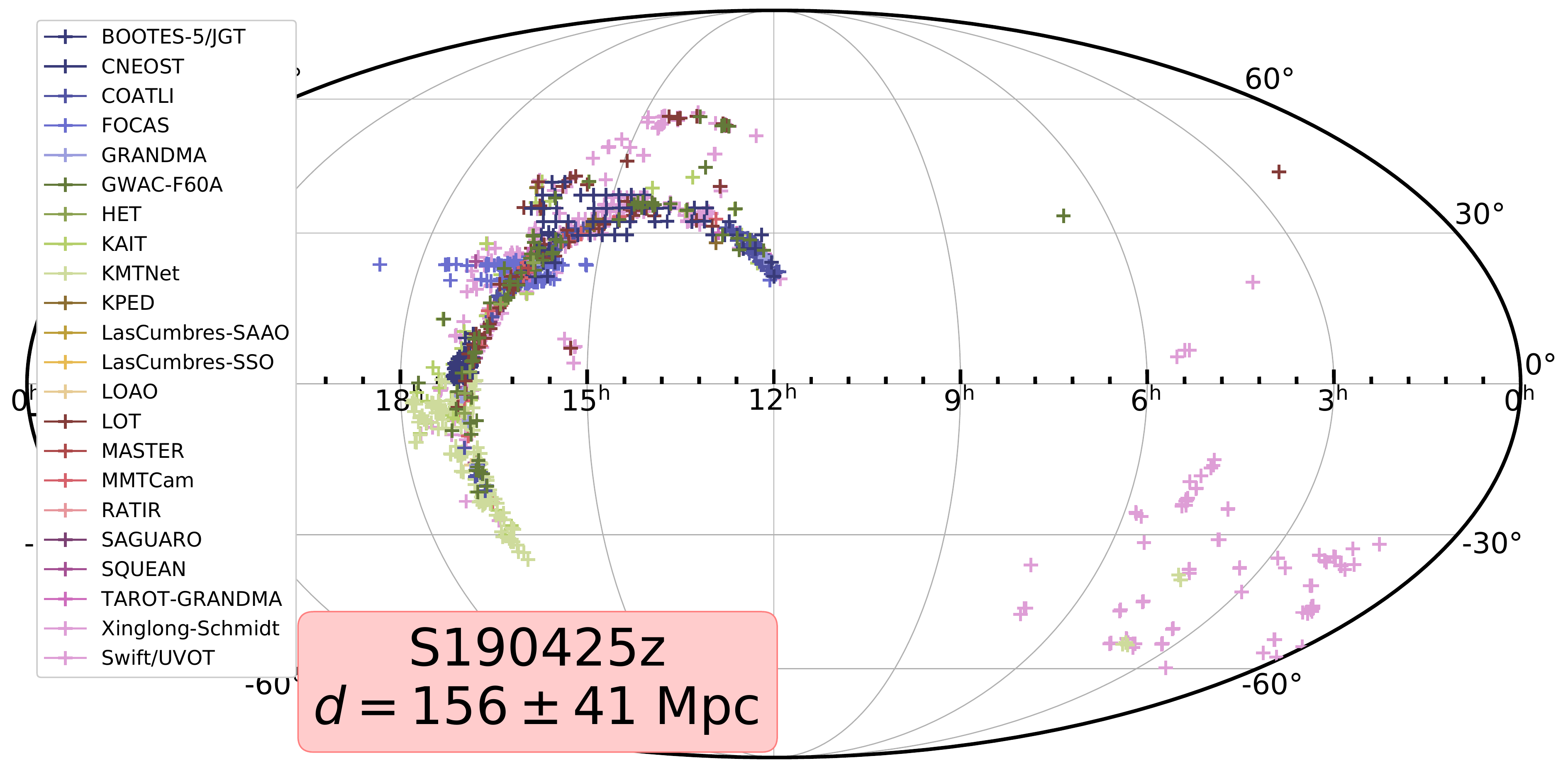}}
\hspace*{0.1in}
{\includegraphics[width=0.33\textwidth]{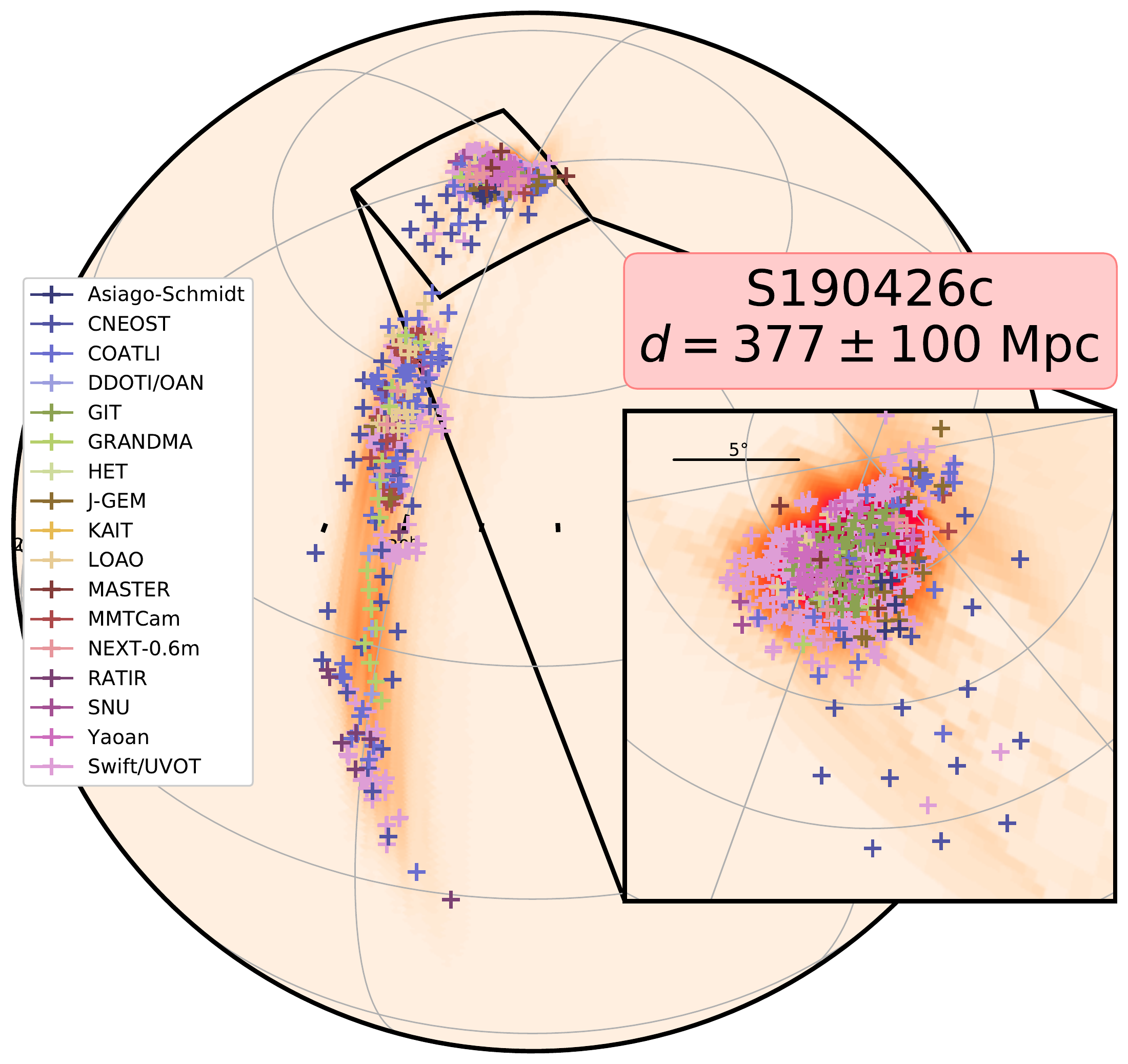}}
\caption{GW localization regions of S190425z (left) and S190426c (right) overlaid with the locations of follow-up observations from our search with MMT, and all publicly reported searches that provided telescope pointing information.  We note that the searches include both galaxy-targeted and wide-field imaging; we do not plot the fields of view of the individual telescopes.
\label{fig:map}}
\end{center}
\end{figure*}

The joint detection of gravitational waves (GW) and electromagnetic (EM) radiation from the binary neutron star (BNS) merger GW170817 was a watershed event. The merger was accompanied by a weak short gamma-ray burst (SGRB), by ultraviolet (UV)/optical/near-infrared (NIR) emission due to a kilonova (during the first month), and by radio, X-ray, and long-term optical emission due to an off-axis jet \citep{2017PhRvL.119p1101A,2017ApJ...848L..12A,2017ApJ...848L..13A}.  GW170817 was localized to a region of about $30$ deg$^2$ and to a distance of $40\pm 8$ kpc, which enabled both galaxy-targeted and wide-field searches to rapidly identify the EM counterpart, within about 11 hr of merger \citep{Arcavi+17,Coulter+17,Lipunov+17,Soares-Santos+17,Tanvir+17,Valenti+17}.

Observing run 3 (O3) of the Advanced Laser Interferometer Gravitational-Wave Observatory (LIGO) and Advanced Virgo (ALV) commenced on 2019 April 1, with a $50\%$ increase in sensitivity compared to Observing Runs 1 and 2. The resulting BNS merger detection distances in O3 are on average about 140 Mpc for LIGO Livingston, 110 Mpc for LIGO Hanford, and 50 Mpc for Virgo.  Given the volumetric merger rate inferred from GW170817, $110-3840$ Gpc$^{-3}$ yr$^{-1}$ \citep{gwtc1}, the expected number of BNS merger detections in the year-long O3 is $\sim 1-20$ (assuming 70\% duty cycle for LIGO). For neutron star (NS)--black hole (BH) mergers the upper bound on the rate based on non-detections in O1 and O2 is $\lesssim 600$ Gpc$^{-3}$ yr$^{-1}$ \citep{gwtc1}; however, given their larger detection volume relative to BNS mergers the observed NS--BH merger rate in O3 may exceed that of BNS mergers.

On 2019 April 25 at 08:18:05.017 UTC ALV detected a GW candidate event, designated S190425z, with a false alarm rate (FAR) of 1 in $7\times 10^4$ yr, a probability of being a BNS merger of $>99\%$, and a luminosity distance of $155\pm 45$ Mpc \citep{gcn24168}. Because the event was detected only by LIGO Livingston and marginally by Virgo (LIGO Hanford was offline at the time of detection), the localization region had an initial area of about $10^4$ deg$^2$ ($90\%$ confidence; \citealt{gcn24168}), which was refined to $7460$ deg$^2$ about 31 hr post merger \citep{gcn24228}. The most up-to-date sky localization region and distance estimate is shown in Figure~\ref{fig:map}.

Soon after, on 2019 April 26 at 15:21:55.337 UTC, ALV detected another GW candidate event, designated S190426c, with an FAR of 1 in 1.7 yr, a probability of containing a NS of $>99\%$, a luminosity distance of $375\pm 108$ Mpc, and an initial localization region of about $1260$ deg$^2$ ($90\%$ confidence; \citealt{gcn24237}), which was refined to $1130$ deg$^2$ about 20 hr post merger \citep{gcn24277}. Given the relatively high FAR, we cannot be certain that the event was astrophysical. However, under the assumption that it was, the latest parameter estimation gives a 60\% probability that the more massive binary component was $>5\ M_\odot$, a 25\% probability that it was $3-5\ M_\odot$, and a 15\% probability that it was $<3\ M_\odot$, suggesting that the NS--BH classification is most likely \citep{gcn24411}. The most up-to-date sky localization region and distance estimate is shown in Figure~\ref{fig:map}.

Here we report our optical follow-up of both events, using the MMT 6.5~m telescope to target galaxies within their localization volumes.  In \S\ref{sec:MMT} we present our MMT observations. In \S\ref{sec:community} we collate searches reported publicly via the GRB Coordinates Network (GCN) circulars to explore a few aspects of the follow-up and announced candidates, as well as our spectroscopic follow-up of two candidates.  In \S\ref{sec:comp} and \S\ref{sec:ag} we compare the results to the kilonova emission of GW170817, to theoretical kilonova models, and to on-axis and slightly off-axis SGRB afterglow models. In \S\ref{sec:high} we collate $\gamma$-ray and X-ray searches and compare to the same SGRB models. We summarize and draw some initial conclusions in \S\ref{sec:conc}.

\section{Galaxy-targeted Follow-up with MMT}
\label{sec:MMT}

\begin{figure*}[!t]
\begin{center}
\scalebox{1.}
\centering
{\includegraphics[width=\textwidth]{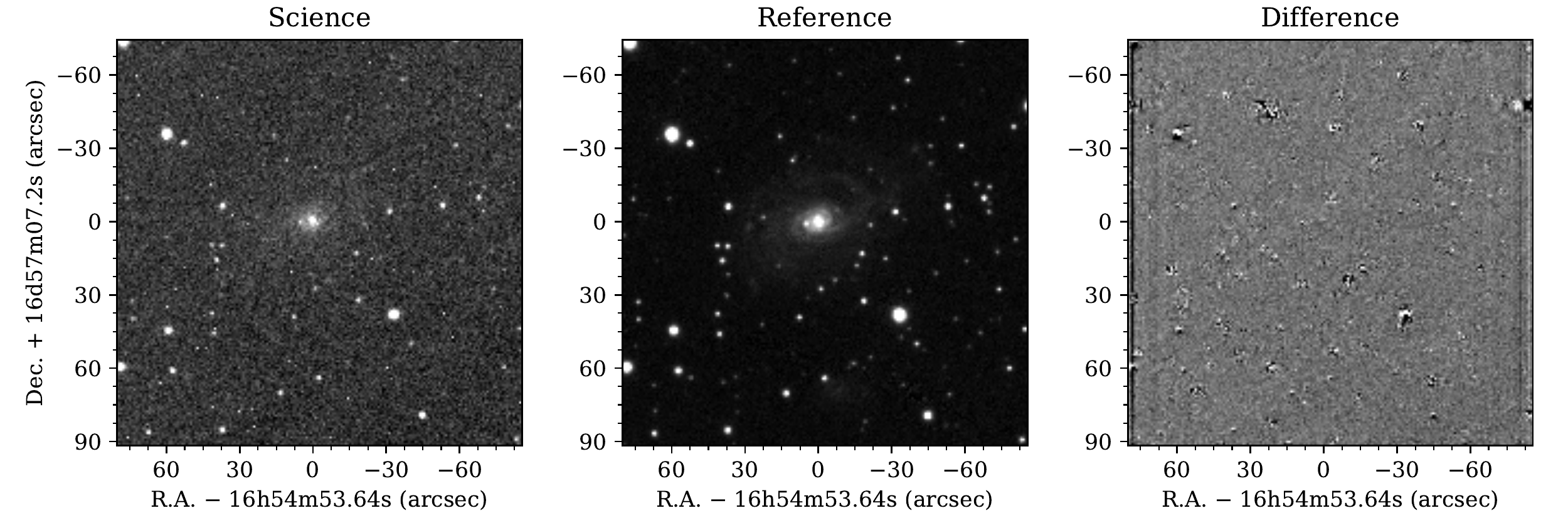}}
\caption{The first galaxy imaged with our MMT program in the field of S190425z (left panel), along with the corresponding reference image from PS1 $3\pi$ (middle panel; \citealt{Chambers_pan-starrs1_2016}), and the resulting subtraction (right panel). The difference image exhibits only astrometric noise and cosmic ray artifacts; no possible counterpart is identified in this image to a limit of $g=22.3$ mag.
\label{fig:mmt}}
\end{center}
\end{figure*}

Since the beginning of O3 we have been using the MMT 6.5~m telescope at Fred L.~Whipple Observatory in Arizona to carry out follow-up observations of galaxies within the localization volumes of GW alerts.  Upon receipt of an alert, our automated software generates a list of galaxies in the Galaxy List for the Advanced Detector Era (GLADE) catalog \citep{dalya_glade:_2018} that are located within the 90\% confidence volume, ranked by probability within the volume.  The software also downloads reference images and catalogs from the Pan-STARRS1 (PS1) $3\pi$ database \citep{Chambers_pan-starrs1_2016} and collates the locations of all previously reported transients and moving objects from the Zwicky Transient Facility Public Survey \citep[via the MARS broker\footnote{\url{https://mars.lco.global}}]{bellm_zwicky_2019,2019PASP..131g8001G,masci_zwicky_2019}, all other public time-domain surveys (via the Transient Name Server\footnote{\url{https://wis-tns.weizmann.ac.il}}), and the Minor Planet Center \citep[via the SkyBoT service;][]{2006ASPC..351..367B}.  A custom data reduction pipeline processes each image as it is read out and performs image subtraction (using PyZOGY;  \citealt{zackay_proper_2016,2017zndo...1043973G}). The reference, science, and subtracted images are then inspected for new transients using a custom web interface based on Flask and JS9 \citep{eric_mandel_2018_1453307}.  An example from our search in the localization region of S190425z is shown in Figure~\ref{fig:mmt}.

For S190425z, we commenced observations using the MMTCam imager on 2019 April 25 at 11:39:23 UT, 3.4 hr post merger, and continued until morning twilight, with our last exposure ending at 12:06:46 UT \citep{gcn24182}. We obtained 30-s $g$-band exposures of 17 galaxies\footnote{Our first GCN circular accidentally omitted four observed galaxies and included one galaxy that was not observed until the next night.}. On the following night, from 08:30:29 to 10:55:00 UT ($24.2-26.6$ hr post merger), we imaged 50 additional galaxies in $i$-band, to minimize moonlight contamination \citep{gcn24244}.  No transient sources were uncovered in these observations to median $3\sigma$ limiting magnitudes of $g=22.0$ and $i=22.5$.  We provide the information for all of the individual galaxies in Table~\ref{tab:mmt}.

For S190426c, we imaged 50 galaxies with 30-s $i$-band exposures on 2019 April 27 at 08:38:51 to 10:15:57 UT ($17.3-18.9$ hr post merger; \citealt{gcn24292}). No transient sources were uncovered in these observations to a median $3\sigma$ limiting magnitude of $i=22.3$ (see Table~\ref{tab:mmt}).

\section{Summary of Community Follow-Up and Our Spectroscopic Follow-up}
\label{sec:community}

Multiple teams reported UV, optical, and NIR follow-up imaging of the sky regions of S190425z and S190426c. In Table~\ref{tab:followup190425z} (S190425z) and Table~\ref{tab:followup190426c} (S190426c) we collate the available information, and summarize the timing of the observations relative to the merger time, the filter(s) used and limiting magnitudes, the sky area covered or number of galaxies targeted, and the relevant GCN circular references.  In Figure~\ref{fig:map} we map the searches that reported their telescope pointing coordinates relative to the GW localization regions. 

\begin{figure}[!t]
\begin{center}
{\includegraphics[width=\columnwidth]{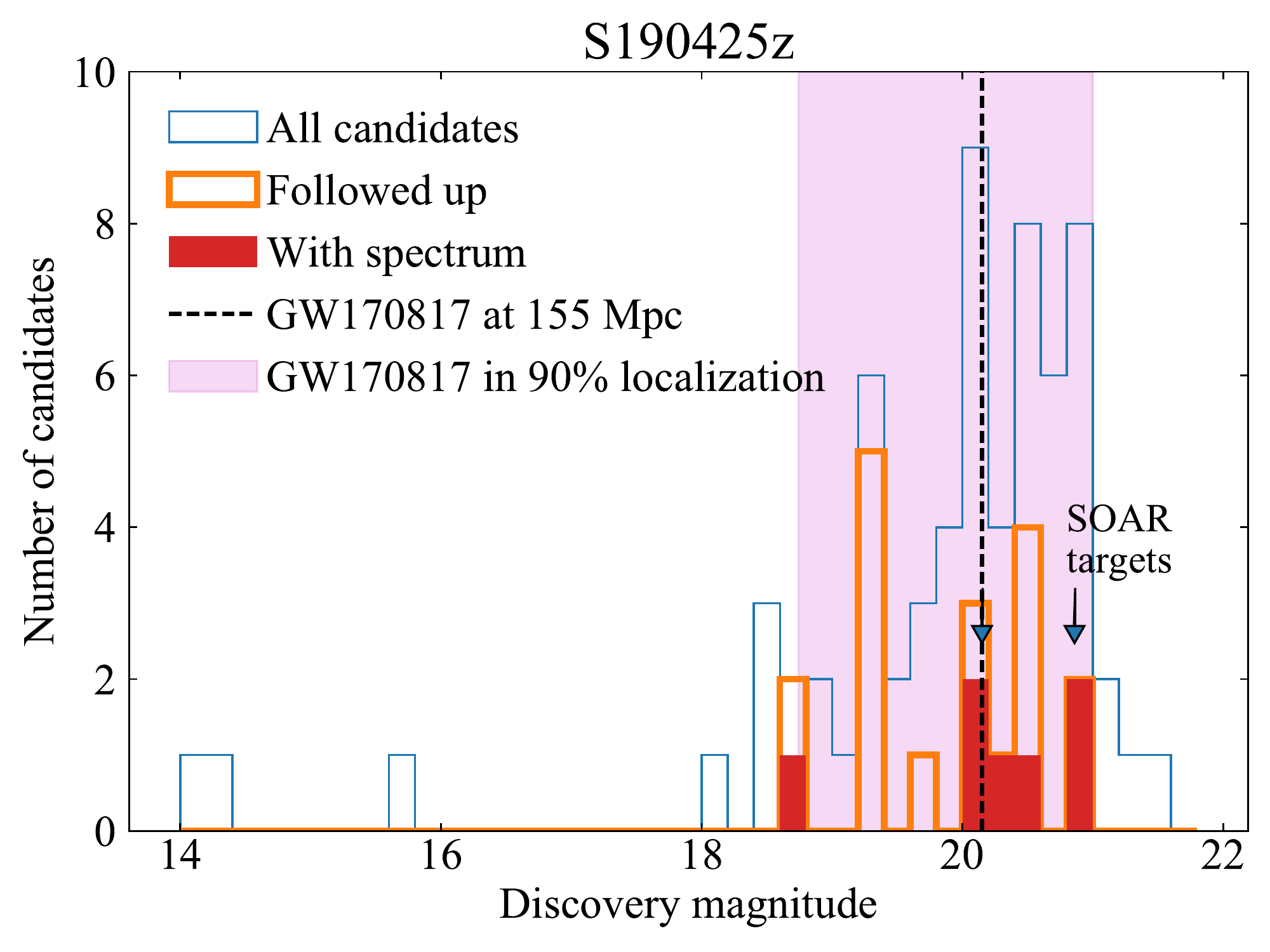}}
{\includegraphics[width=\columnwidth]{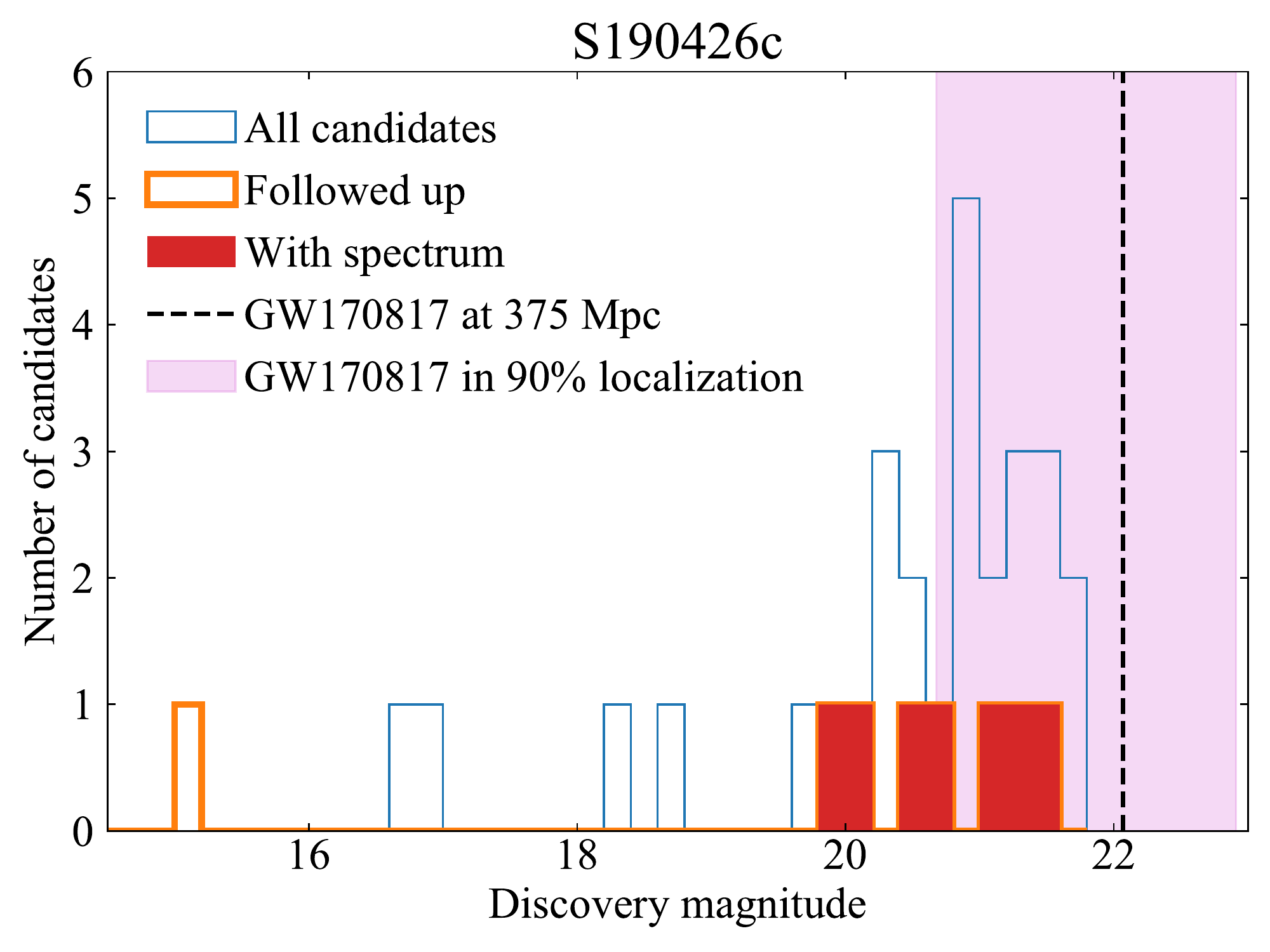}}
\caption{Top panel: distribution of magnitudes (in $u,g,r,i,o$ bands) for all reported candidates in the S190425z localization region (blue), those with any optical/NIR imaging follow-up (orange), and those with spectroscopic follow-up (red); the latter were all found to be normal supernovae. The target classified through our SOAR program are marked with arrows \citep{gcn24321}.  The vertical bar indicates the optical peak brightness of GW170817 for the 90\% distance range of S190425z. Bottom panel: the same plot but for S190426c.
\label{fig:candmags}}
\end{center}
\end{figure}

\subsection{S190425z}

\begin{figure}[!t]
\begin{center}
{\includegraphics[width=\columnwidth]{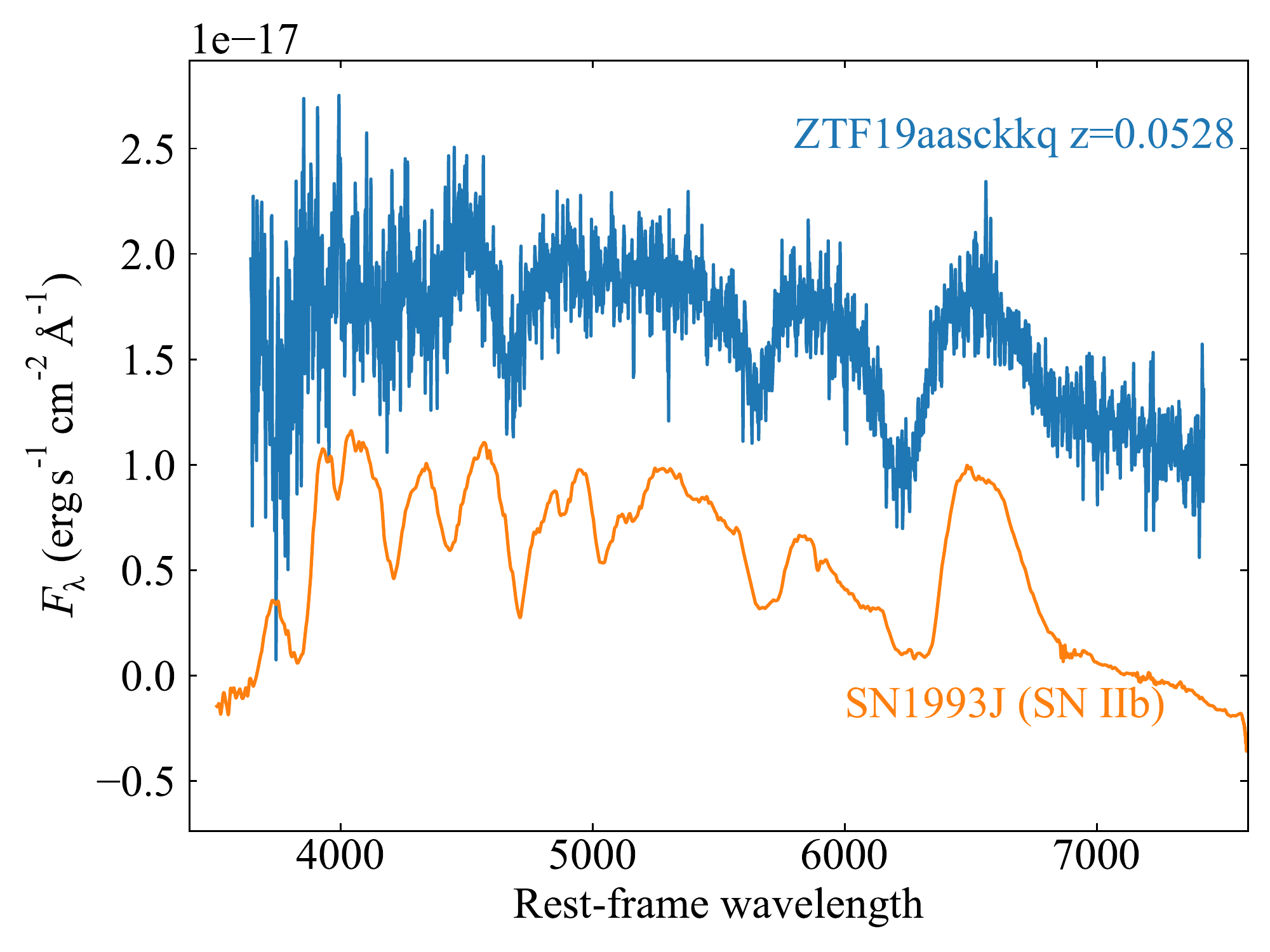}}
{\includegraphics[width=\columnwidth]{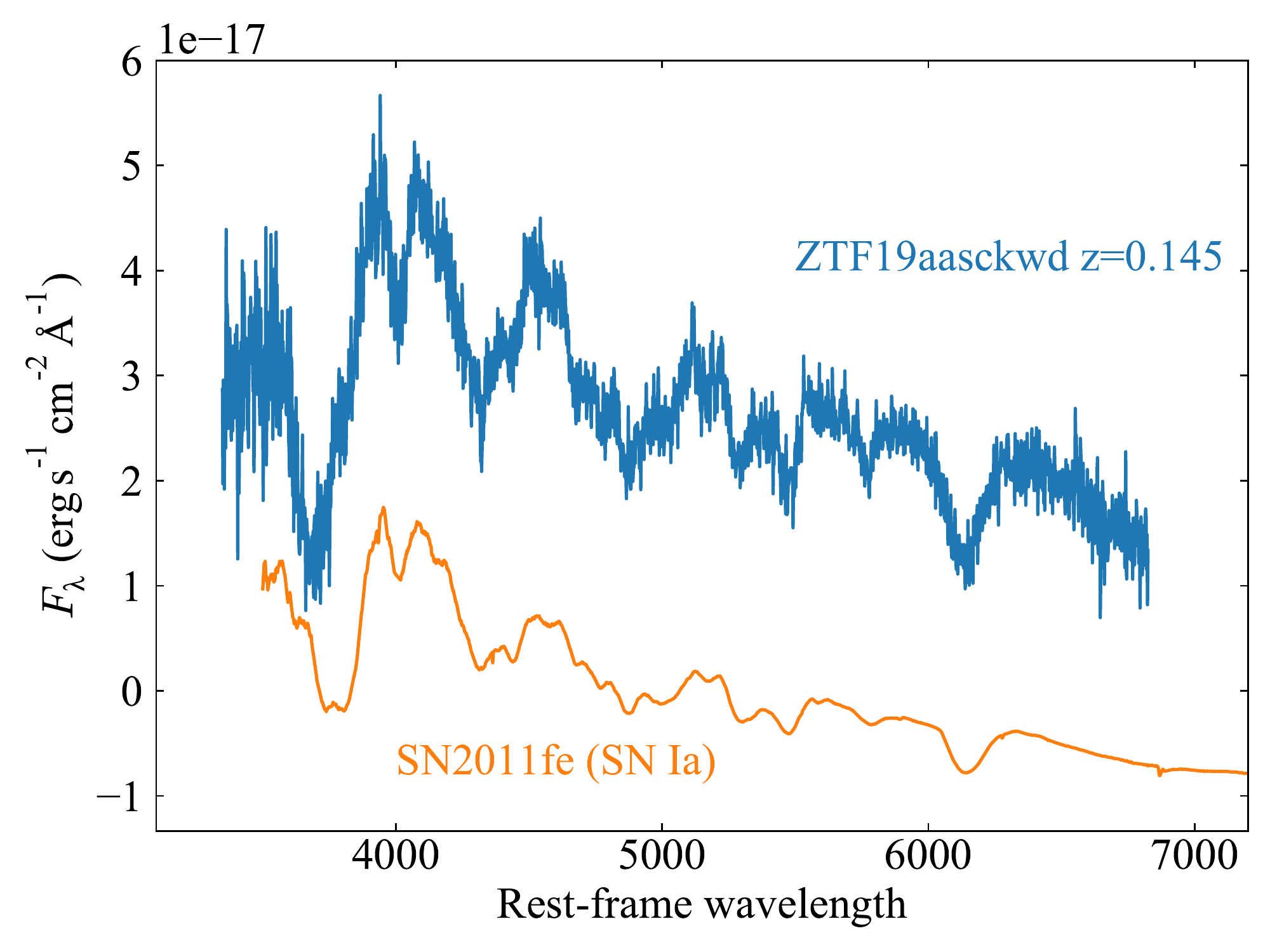}}
\caption{Spectroscopic classification using SOAR of optical transients in the S190425z localization region \citep{gcn24321}. Top panel: ZTF19aasckkq was selected based on a host spectroscopic redshift of $z=0.0528$ and transient absolute magnitude of $\approx -16.3$.  We classify this event as a SN~IIb, as shown by comparison to SN\,1993J. Bottom panel: ZTf19aasckwd was selected based on an apparent magnitude consistent with GW170817 at 155 Mpc. We classify this event as a SN~Ia at $z=0.145$, as shown by comparison to SN\,2011fe.
\label{fig:soar}}
\end{center}
\end{figure}

In all, 24 telescopes were reported to follow up S190425z, whose 90\% confidence localization volume is about $8\times 10^6$ Mpc$^3$. The galaxy-targeted searches observed a combined total of 418 galaxies in this volume, corresponding to about $1\%$ of the total number of galaxies in the GLADE catalog within the volume. Most searches observed galaxies more luminous than \mbox{$M_B\approx -19$ mag} (373 galaxies). Integrating the galaxy luminosity function down to this limit indicates $2.5\times 10^4$ galaxies within the localization volume, confirming that the GLADE catalog is effectively complete at this luminosity. The fraction of observed bright galaxies was thus about $1.5\%$. To further quantify the effective coverage of the galaxy-targeted searches, we consider only the galaxies that were imaged to a sufficient depth to detect a GW170817-like kilonova ($M\approx -16$ mag; \citealt{Villar+17a}) at the distance of each galaxy. We find that 304 out of the 373 galaxies satisfy this criterion, leading to an effective coverage of about $1.2$\%.  In the UV, {\it Swift}/Ultra-Violet Optical Telescope (UVOT) observed 389 galaxies that are both in the GLADE catalog and have a redshift that would make it possible to detect a GW170817-like kilonova at the depth of the observations; this corresponds to an effective coverage of about 1.5\%.

Similarly, for the wide-field searches we determine the effective fractional volume coverage using the distance to which each search would have detected a GW170817-like kilonova at the reported limiting magnitude, and combine this effective distance with the reported areal coverage.  We find that most of the wide-field searches had an effective fractional volume coverage of about $0-8\%$, while ZTF had a fractional volume coverage of about $40\%$; the values of zero correspond to searches that reported limiting magnitudes too shallow to have detected a GW170817-like kilonova at the lower distance limit of $\approx 115$ Mpc.

Naturally, the fractional coverage of the galaxy-targeted and wide-field searches would be smaller (larger) for a dimmer (brighter) counterpart than in GW170817.  We also note that our calculation is simplified, and likely errs on the side of being too optimistic.  For example, we are not taking into account variations in Galactic extinction, moon illumination, and other differential observational effects that would generally serve to reduce the efficiency of the searches. On the other hand, other groups may have conducted follow-up campaigns that have not (yet) been publicly reported, which may increase the overall efficiency of the community effort. Lastly, the numbers in Tables~\ref{tab:followup190425z} and \ref{tab:followup190426c} are uncertain due to possible human errors in real-time GCN composition (as ours had), but we assume these uncertainties are small compared to the uncertainty in the kilonova models.

The various searches returned 69 candidate optical counterparts, reported by ZTF, ATLAS, Pan-STARRS, {\it Swift}/UVOT, and \textit{Gaia}, with candidates ranging in brightness from about 14 to 21.5 mag \citep{gcn24191,gcn24197,gcn24210,gcn24296,gcn24311,gcn24345,gcn24362}; see Figure~\ref{fig:candmags} for the brightness distribution. Of these, 18 candidates were followed up with at least one targeted observation, including seven events that were spectroscopically classified. 

Two of the classifications were obtained through our spectroscopic follow-up program using the 4.1~m SOAR telescope \citep{gcn24321}. ZTF19aasckkq was selected based on a known spectroscopic redshift of $z=0.0528$ for its host galaxy \citep{gcn24311}, which is within the $2\sigma$ contour of the the GW localization distance. The absolute magnitude of the source at this distance was $-16.3$\,mag, comparable to GW170817 at peak. We obtained a 1900 s exposure beginning at 2019 April 28 05:13:50 UT using the Goodman High Throughput Spectrograph \citep{clemens2004} with the 400 lines/mm grating and a central wavelength of 5750\,\AA. The data were processed and the one-dimensional spectrum extracted using a custom \texttt{Iraf} pipeline (for details, see \citealt{Margutti2019}). The spectrum shows broad \ion{H}{1} lines at the redshift of the host galaxy, as well as a strong absorption consistent with \ion{He}{1} $\lambda5876$. Classification using the Supernova Identification code \citep[SNID;][]{Blondin2007} indicates that this transient is likely a young Type~IIb supernova (SN~IIb; Figure~\ref{fig:soar}). Our second SOAR target was ZTf19aasckwd. This event did not have a reported redshift, but the apparent magnitude of 20.15 was a good match to the brightness of GW170817 at $\sim 155$ Mpc. Moreover, a search of the coordinates in PS1 $3\pi$ data \citep{Flewelling2016} showed a likely host galaxy. We obtained a 1500 s exposure in the same setting as above, beginning at 2019 April 28 04:41:03 UT. The spectrum revealed a SN~Ia at $z=0.145$ (Figure~\ref{fig:soar}). Both spectra are publicly available via the Transient Name Server (ZTF19aasckkq = SN~2019eff, ZTf19aasckwd = SN~2019eib). 

The other five candidates classified by the community also turned out to be normal SNe: one SN~Ia, three SNe~II, and one SN~Ib/c \citep{gcn24200,gcn24204,gcn24220,gcn24205,gcn24208,gcn24209,gcn24211,gcn24214,gcn24215,gcn24217,gcn24221,gcn24230,gcn24233,gcn24252,gcn24295,gcn24358,gcn24269,gcn24311}. Additionally, a UV candidate uncovered by \textit{Swift}/UVOT \citep{gcn24296} was shown to be an M dwarf flare \citep{gcn24326,gcn24337}.  No NIR candidates were announced. 

As shown in Figure~\ref{fig:candmags}, the bulk of reported candidates overlapped the optical brightness of a GW170817-like kilonova in the 90\% confidence distance range ($\approx 19-21$ mag).  We further find that the subset of events followed up photometrically and/or spectroscopically similarly span the same magnitude range.  In terms of the choice of targets for spectroscopic follow-up, four of the seven classified transients (and four of the 11 transients with only photometric follow-up) were selected based on probable associations with galaxies that have secure distance measurements compatible with the GW distance \citep{gcn24191,gcn24210,gcn24311}.

Conversely, six of of the 11 sources with only photometric follow-up were announced as apparently host-less (or ``orphan'') transients \citep{gcn24197,gcn24311}. None of these sources were recovered in follow-up imaging \citep{gcn24202,gcn24211,gcn24343}, suggesting that they were potential image artifacts or due to stellar variability.  Follow-up of host-less transients therefore appears to be a somewhat risky strategy, although we note that some mergers may occur at large offsets from their hosts: based on the distribution of SGRB offsets \citep{FongBerger13}, about 10\% of BNS and/or NS--BH mergers may have offsets of tens of arcseconds from their hosts at the ALV detection distances.

\subsection{S190426c}

In all, 21 telescopes were reported to follow up S190426c, whose 90\% confidence localization volume is about $2\times 10^7$ Mpc$^3$. The searches that targeted individual galaxies observed a combined total of 378 galaxies in this volume, corresponding to about $3.5\%$ of the total number of galaxies in the GLADE catalog within the volume; however, the GLADE catalog is highly incomplete at the distance of S190426c. Instead, integrating the galaxy luminosity function at $M_B\lesssim -19$ mag, we find about $3.1\times10^4$ galaxies within the volume. Thus, the galaxy-targeted searches covered about $1.2\%$ of the galaxies. Comparing the depth of the searches to the expected brightness of a GW170817-like kilonova at the distance of S190426c ($21-23$ mag), we find an effective fractional coverage of about $0.1\%$.

For the wide-field searches we determine the effective fractional volume coverage in the same manner as for S190425z. We find that most of the wide-field searches had an effective fractional volume of $\approx 0\%$ because they did not reach sufficient depth to detect a GW170817-like kilonova even at the lower bound of the distance range.  However, ZTF and DECam had effective volume coverages of about $55\%$ and $8\%$ \citep{Goldstein19}, respectively\footnote{The DECam observations covered a southern probability region that was mostly eliminated in the revised localization map released after the DECam observations occurred.}.

In total, 30 candidate optical counterparts were reported by Las Cumbres Observatory, DECam, ZTF, GRAWITA, and \textit{Gaia}, ranging from about 15--21.5 mag \citep{gcn24251,gcn24268,gcn24283,gcn24331,gcn24340,gcn24344}; see Figure~\ref{fig:candmags} for the brightness distribution. Of these, eight were observed spectroscopically, leading to classifications of three SNe~Ia, one SN~II, one broad-lined SN~Ic, and one Galactic cataclysmic variable \citep{gcn24317,gcn24359,gcn24368}. The other two transients were not detected in their follow-up spectra \citep{gcn24275,gcn24430}. Only one additional candidate was followed up with imaging rather than spectroscopy. About half of the candidate counterparts and the subset classified spectroscopically are brighter than a GW170817-like kilonova in the 90\% distance range ($\approx 21-23$ mag; Figure \ref{fig:candmags}).

\begin{figure*}[!t]
\begin{center}
\hspace*{-0.1in}
\scalebox{1.}
{\includegraphics[width=0.495\textwidth]{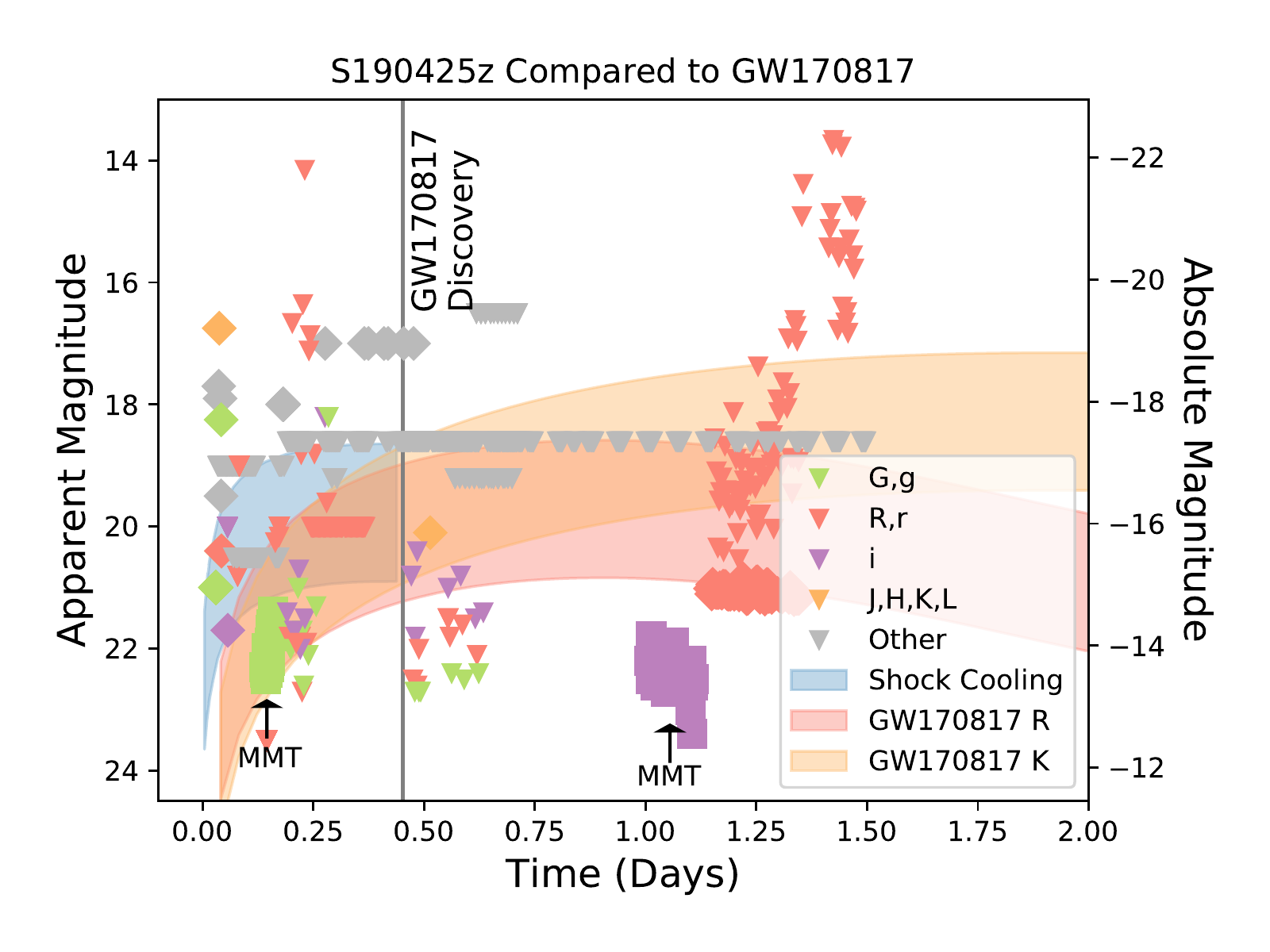}}
{\includegraphics[width=0.495\textwidth]{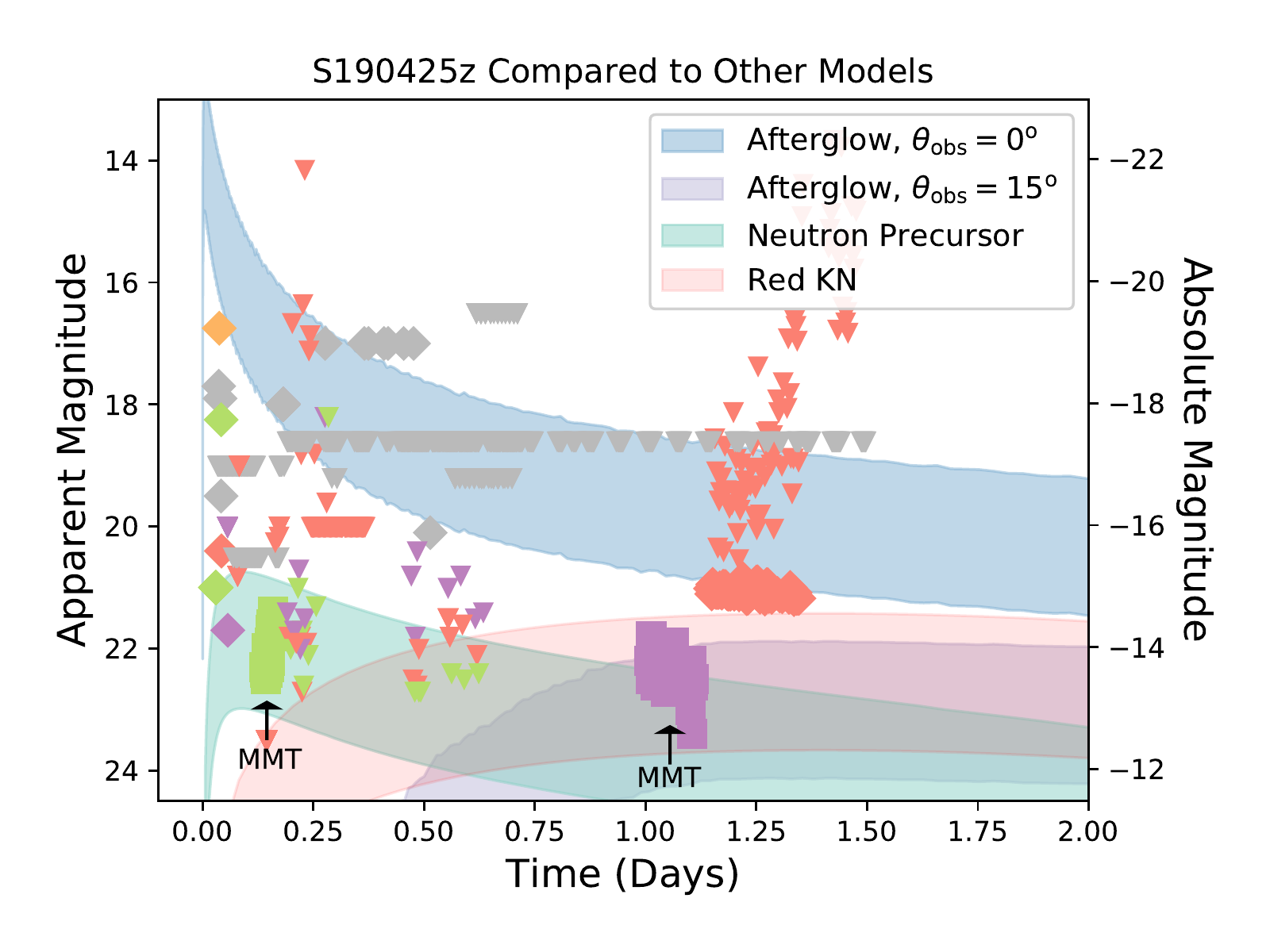}}
{\includegraphics[width=0.495\textwidth]{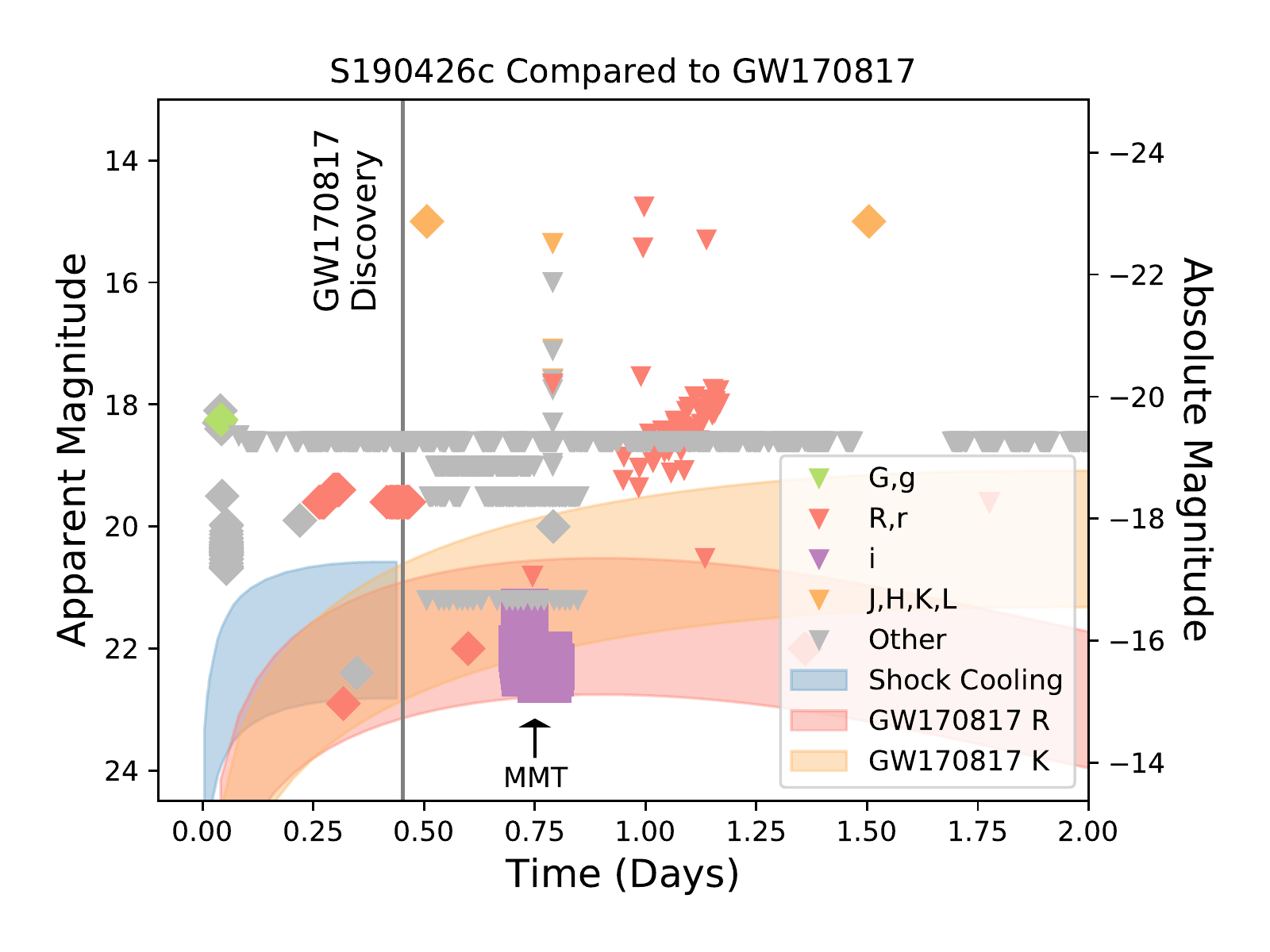}}
{\includegraphics[width=0.495\textwidth]{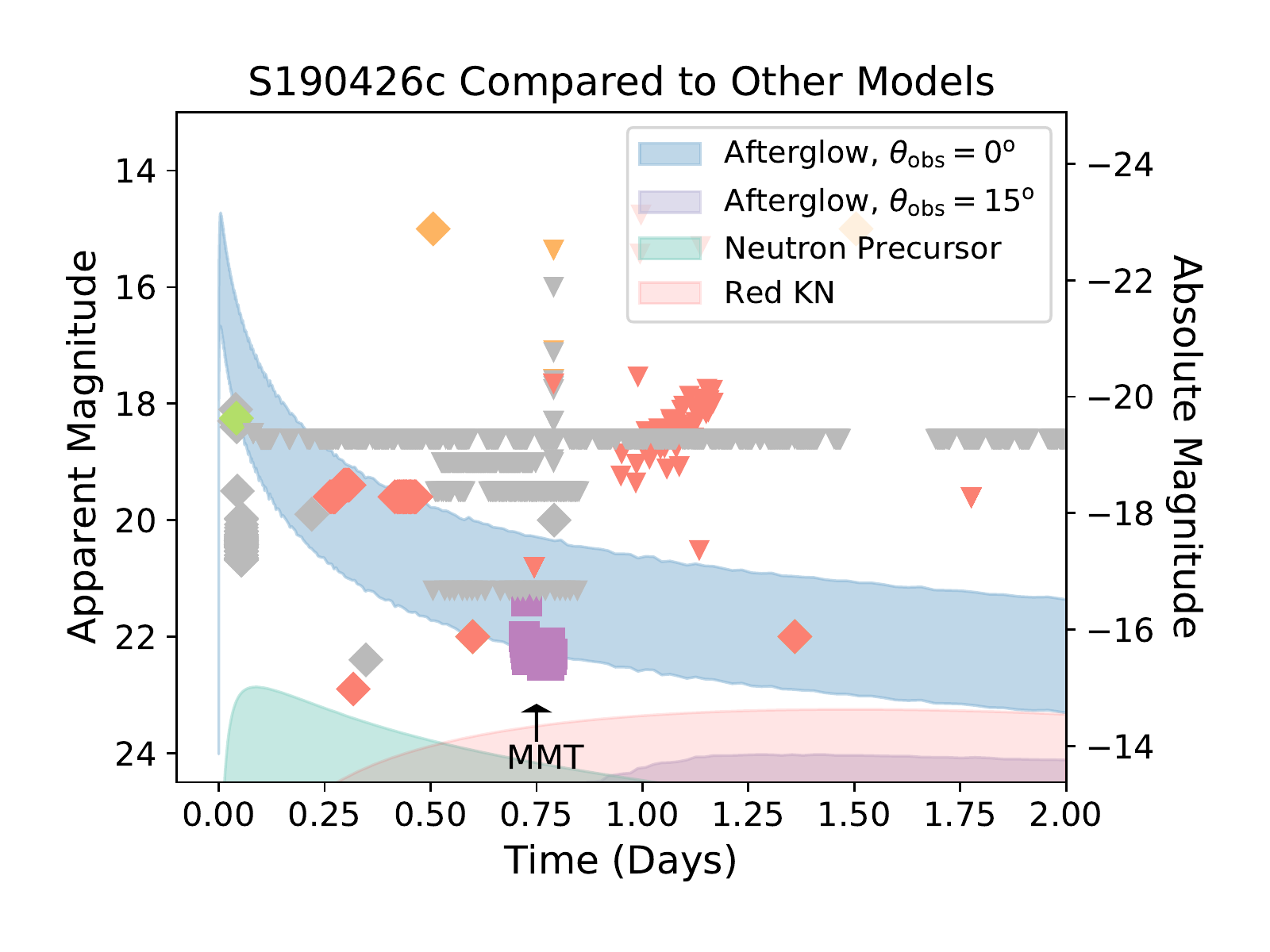}}
\caption{Top row: limiting magnitudes as a function of time post merger for UV/optical/NIR searches in the localization region of S190425z (diamonds represent wide-field searches; triangles represent galaxy-targeted searches; colors correspond to different filters).  Left panels: a comparison to the kilonova in GW170817 scaled to the 90\% confidence distance range of S190425z in $r$-band (red) and $K$-band (orange) based on the model fits from \citet{Villar+17a}. Additionally shown is a model of shock cooling emission for the early emission from GW170817 (blue; \citealt{PiroKollmeier18}).  The vertical line marks the time when the optical counterpart of GW170817 was first detected. Right panels: comparison to on-axis (blue) and slightly off-axis (purple) afterglow models (based on SGRBs; \citealt{Fong+15}), to a lanthanide-rich kilonova with an ejecta mass of $0.01\ M_\odot$ (red), and to a neutron precursor (green; \citealt{Metzger+15}).  Bottom row: same as top row, but for observations of S190426c, with the models scaled to its 90\% confidence distance range.
\label{fig:comp}}
\end{center}
\end{figure*}

\section{Comparison to GW170817 and Kilonova Models}
\label{sec:comp}

In Figure~\ref{fig:comp} (left panel) we compare the limiting magnitudes of the various searches (galaxy-targeted and wide-field) to the model optical/NIR light curves of GW170817 (from \citealt{Villar+17a}\footnote{These models are based on data obtained by \cite{andreoni2017follow}, \cite{Arcavi+17}, \cite{Coulter+17}, \cite{cowperthwaite2017electromagnetic}, \cite{diaz2017observations}, \cite{drout2017light}, \cite{evans2017swift}, \cite{hu2017optical}, \cite{kasliwal2017illuminating}, \cite{Lipunov+17}, \cite{pian2017spectroscopic}, \cite{pozanenko2018grb}, \cite{shappee2017early}, \cite{smartt2017kilonova}, \cite{troja2017x}, \cite{utsumi2017j}, and \cite{Valenti+17}.}) shifted to the 90\% distance ranges of S190425z and S190426c.  We also show the shock cooling model of \citet{PiroKollmeier18}, as it predicts potentially brighter emission in the first few hours post merger (which were missed in the case of GW170817).  For both events some of the searches reached sufficient depth to detect a GW170817-like kilonova, although this was more challenging for S190426c.  

In the right panel of Figure~\ref{fig:comp} we compare the searches to several other models of early optical/NIR emission.  We consider a kilonova that lacks the blue (lanthanide-poor) component and has an ejecta mass of $0.01\ M_\odot$ for the red (lanthanide-rich) component (i.e., about 4 times lower than in GW170817); this model represents a more pessimistic possibility, but one that is supported by binary merger simulations (e.g., \citealt{Hotokezaka+11}).  Both models were generated with MOSFiT \citep{Guillochon+17b}. We also show a model for a blue precursor powered by the decay of free neutrons from the shock-heated interface between the NSs \citep{Metzger+15}. In the context of these models, which peak at $\approx 22$ mag for S190425z and $\approx 24$ mag for S190426c, we find that few (if any) observations reached sufficient depth to place meaningful constraints.

\section{Comparison to On-axis and Off-axis SGRB Afterglows}
\label{sec:ag}

Another source of early UV/optical/NIR emission is an on-axis or slightly off-axis relativistic jet, as observed in SGRBs \citep{Berger2014}.  In the absence of information about the binary inclination from GW data we cannot directly assess the viewing angle of a potential jet, but we note that based on jet opening angle measurements in SGRBs \citep{Fong+15} we expect at most a few percent of GW mergers to exhibit on-axis jets \citep{MetzgerBerger12}.  Conversely, for substantial off-axis angles (as was the case for GW170817 with $\theta_{\rm obs}\approx 30^{\circ}$; \citealt{Alexander+17,Alexander+18,Margutti+17,Margutti+18}) the optical emission is significantly delayed and exceedingly dim, making this scenario irrelevant for the rapid searches considered here.  

We therefore consider two afterglow models: on-axis ($\theta_{\rm obs}=0^{\circ}$) and slightly off-axis ($\theta_{\rm obs}=15^{\circ}$). We generate light curves using the \texttt{BOXFIT} code (v2; \citealt{vanEerten+11}) for ``top hat'' jets\footnote{For the small viewing angles considered here, a more complex jet structure (as was inferred for GW170817; \citealt{Alexander+18,Margutti+18,WuMacFadyen18}) will make little difference.} using median values for cosmological SGRBs \citep{Fong+15}: jet opening angle of $\theta_j=10^{\circ}$, isotropic kinetic energy of $E_{\rm K,iso}=2\times 10^{51}$ erg, circumburst density of $n=4\times 10^{-3}$ cm$^{-3}$, electron energy power-law index $p=2.4$, and fractional post-shock energies in the relativistic electrons and magnetic fields of $\epsilon_E=0.1$ and $\epsilon_B=0.01$, respectively. We note that this model is comparable to the inferred properties of the relativistic jet in GW170817.

The resulting model light curves, scaled to the distances of S190425z and S190426c, are shown in Figure~\ref{fig:comp} (right panels).  We find that for S190425z, the on-axis afterglow model remains brighter than about 20 mag for the first day, exceeding the expected brightness of a GW170817-like kilonova.  A substantial fraction of the searches reached sufficient depth to detect such an on-axis jet.  In the case of S190426c, however, such an on-axis afterglow would have declined below 20 mag within about 0.3 days, although it would have still been detectable by at least some of the searches in the first day.  We stress again that the probability of an on-axis merger is at most a few percent, so even in the case when the observations are sufficiently deep and cover the entire GW localization region, such a detection is unlikely.  For the slightly off-axis afterglow model, the peak brightness is about 23 mag for S190425z and about 25 mag for S190426c.  These brightness levels are well below the limiting magnitudes of the majority of follow-up observations, indicating the challenge of detecting off-axis afterglows at these distances.

\begin{figure}[!t]
\begin{center}
\hspace*{-0.15in}
\scalebox{1.}
{\includegraphics[width=0.5\textwidth]{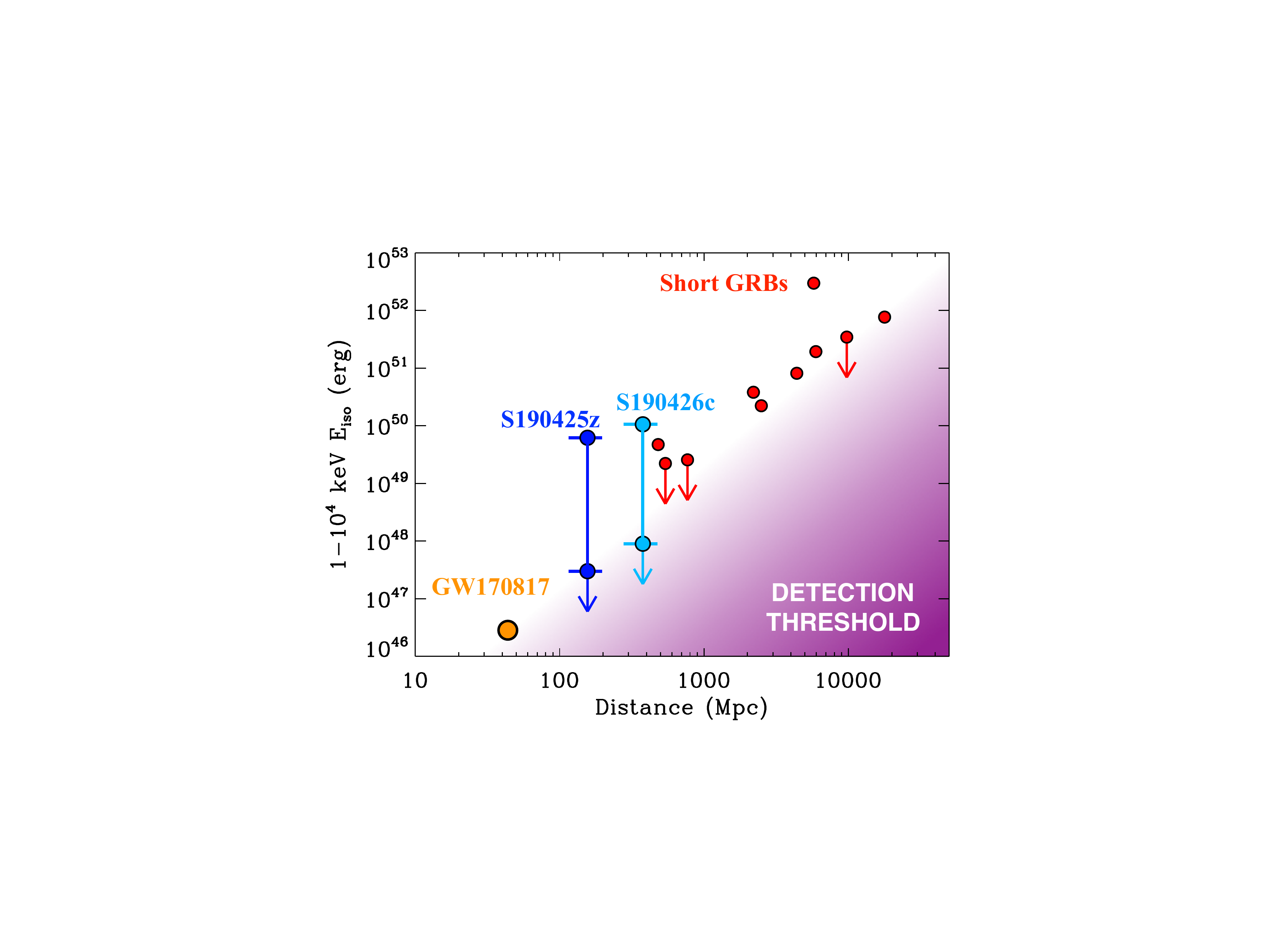}}
\caption{Limits on the isotropic prompt $\gamma$-ray energy release from S190425z (dark blue) and S190426c (light blue) as constrained by Fermi-GBM, INTEGRAL, and Konus-\textit{Wind} observations (\citealt{gcn24185,gcn24248,gcn24417,gcn24418}; V.~Savchenko et al.\ 2019, in preparation). The range of luminosity limits reflects the assumed spectral model used for the flux calibration. For S190425z (S190426c) Fermi-GBM  covered about $50\%$ ($100\%$) of the initial GW probability map, while Konus-\textit{Wind} covered the entire sky. These observations can rule out the most energetic on-axis cosmological SGRBs (red circles; \citealt{2017ApJ...848L..13A}). Also shown for completeness is GRB170817 associated with GW170817 \citep{2017ApJ...848L..13A}.
\label{fig:prompt}}
\end{center}
\end{figure}

\section{Gamma-Ray and X-Ray Follow-up}
\label{sec:high}

\begin{figure*}[!t]
\begin{center}
\hspace*{-0.1in}
\scalebox{1.}
{\includegraphics[width=0.48\textwidth]{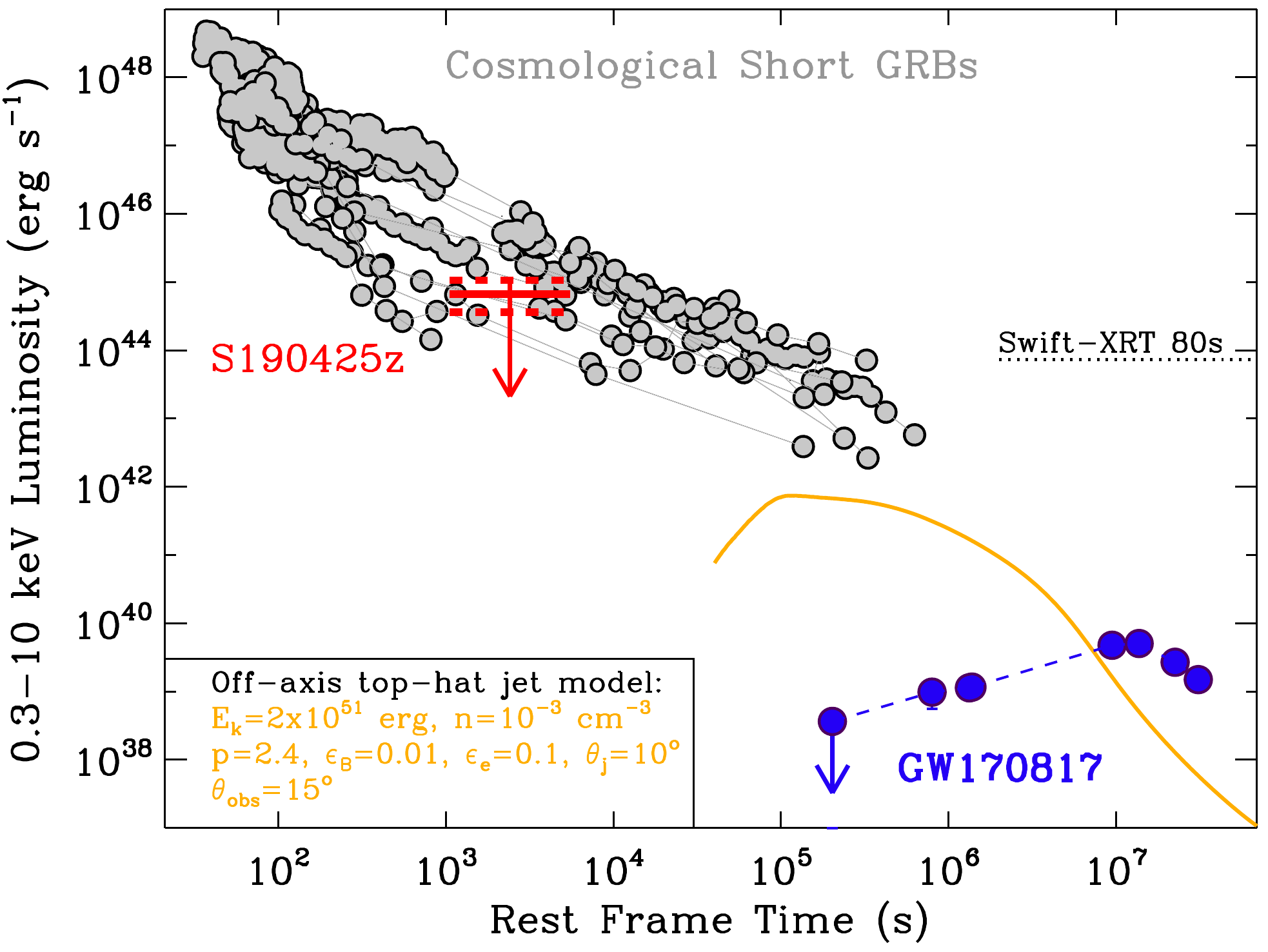}}
\hspace{0.1in}
{\includegraphics[width=0.48\textwidth]{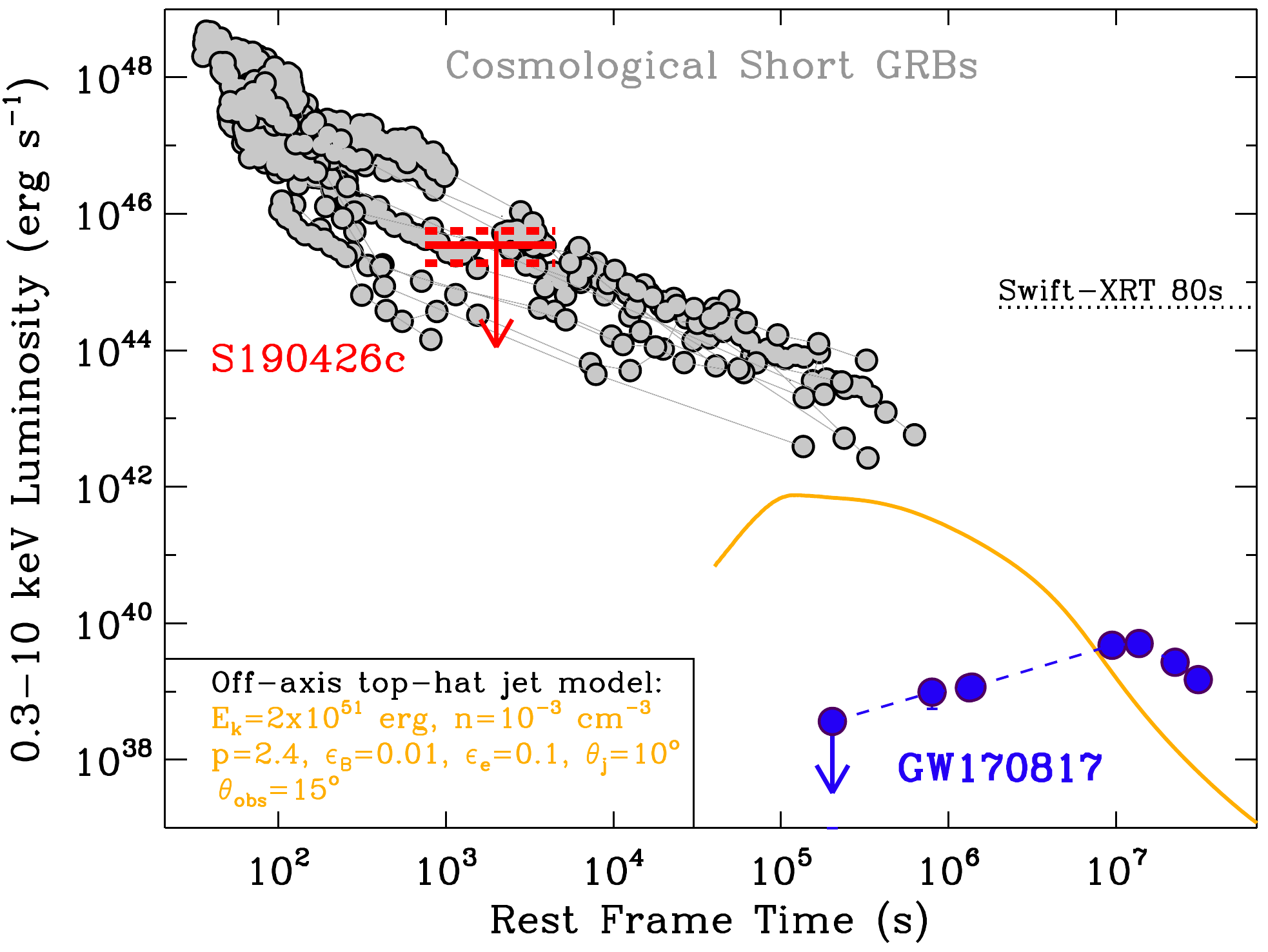}}
\caption{MAXI X-ray observations (red) of S190425z (left panel, initial probability map coverage of $81\%$; \citealt{gcn24259}), and S190426c (right panel, initial probability map coverage of $76\%$; \citealt{gcn24177}).  The limits rule out luminous cosmological SGRB X-ray afterglows (gray points). An off-axis jet as described in \S\ref{sec:ag} (orange line) cannot be ruled out. Also shown are the luminosity limits for \emph{Swift}/XRT observations with exposures of about $80$ s (dotted line; e.g., \citealt{gcn24273,gcn24305,gcn24353}), and the X-ray light curve of GW170817 from \cite{Margutti+18}, \cite{Alexander+18}, and \cite{Hajela2019}.
\label{fig:MAXI}}
\end{center}
\end{figure*}

Multiple $\gamma$-ray and X-ray space missions reacted to the GW alerts and/or were observing parts of the relevant sky area at the time of merger, including the Monitor of All-sky X-ray Image (MAXI), the Neil Gehrels \emph{Swift} Observatory, the INTErnational Gamma-Ray Astrophysics Laboratory (INTEGRAL), the High-Altitude Water Cherenkov Observatory (HAWC), the Fermi Gamma-ray Space Telescope, the Astro$‐$Rivelatore Gamma a Immagini Leggero (AGILE), the Hard X-ray Modulation Telescope (Insight-HXMT) and the CALorimetric Electron Telescope (CALET). No high-energy counterpart was identified with high statistical significance for either S190425z or S190426c \citep{gcn24170, gcn24173,gcn24266, gcn24177, gcn24180, gcn24181, gcn24184, gcn24185, gcn24186, gcn24213, gcn24369, gcn24218, gcn24342, gcn24235, gcn24246, gcn24248, gcn24255, gcn24259, gcn24273, gcn24276, gcn24280}. The fractional localization coverage at the time of the GW detection varies widely, from a few percent to nearly $100\%$.   

For S190425z there are four measurements of interest: (i) \citet{gcn24185} report a Fermi-GBM $3\sigma$ flux limit in the range $F_{10-1000\ \rm{keV}}<(0.1-3)\times 10^{-6}\ \rm{erg\ s^{-1}\ cm^{-2}}$ (depending on the assumed spectral model) for observations obtained at $\pm 30$ s relative to the merger time using a $1\,\rm{s}$ integration time. This corresponds to a luminosity limit of $L_{1-10^{4}\ \rm{keV}}<(0.03-6)\times 10^{49}\,\rm{erg\ s^{-1}}$ at $155\ \rm{Mpc}$. These observations covered $51\%$ of the initial probability map. (ii) Konus-\textit{Wind} was observing the entire sky at the time of the GW trigger. \cite{gcn24417} reported a flux limit of $2.7\times 10^{-7}\ \rm{erg\ s^{-1}\ cm^{-2}}$ (20--1500~keV) for a SGRB-like spectrum. (iii) \citet{gcn24169} and \citet{gcn24178} reported the presence of a possible excess of $\gamma$-ray emission with limited significance in INTEGRAL data acquired $\sim6$ s after the GW detection. However, this excess is most
probably due to background fluctuations. Under this preferred assumption,
V.~Savchenko et al.\ \citetext{2019, in preparation} estimate a typical $3\sigma$ upper limit on the 75-2000 keV fluence
within 50\% GW probability containment region of $(2-6)\times10^{-7}\ \rm{erg\ cm^{-2}}$ depending on the sky location and for a burst lasting less than 1~s with a typical SGRB spectrum. (iv) In the time interval of 1054--5520~s post merger MAXI observed an area of the sky corresponding to $81\%$ of the  probability map, with a $1\sigma$ flux limit of $F_{4-10\ \rm{keV}}<2\times10^{-10}\ \rm{erg\ cm^{-2}\ s^{-1}}$ corresponding to $L_{4-10\ \rm{keV}}<6\times10^{44}\ \rm{erg\ s^{-1}}$ \citep{gcn24177}. 

Similarly, for S190426c there are four measurements of interest: (i) \citet{gcn24248} reported a Fermi-GBM $3\sigma$ flux limit in the range $F_{10-1000\ \rm{keV}}<(1-9)\times 10^{-7}\ \rm{erg\ s^{-1}\ cm^{-2}}$ at $\pm 30$~s relative to merger for a 1~s integration time, corresponding to $L_{1-10^{4}\ \rm{keV}}<(0.1-10)\times 10^{49}\ \rm{erg\ s^{-1}}$ at 377~Mpc. These observations covered about $100\%$ of the initial probability region.  (ii) From Konus-\textit{Wind} observations covering the entire sky, \cite{gcn24418} reported a flux limit of $7.3\times 10^{-7}\ \rm{erg\ s^{-1}\ cm^{-2}}$ (20--1500~keV) for a SGRB-like spectrum. (iii) \emph{Swift}/BAT observations covering $95\%$ of the initial probability region obtained at $\pm 100$ s relative to merger indicate $F_{15-350\ \rm{keV}}<10^{-6}\ \rm{erg\ cm^{-2}\ s^{-1}}$ for a 1~s integration time, corresponding to $L_{15-350\ \rm{keV}}<2\times10^{49}\ \rm{erg\ s^{-1}}$ \citep{gcn24255}. (iv) MAXI covered $76\%$ of the probability region at 750--4488~s post merger to a $1\sigma$ limit of $F_{4-10\ \rm{keV}}<2\times10^{-10}\ \rm{erg\ cm^{-2}\ s^{-1}}$, corresponding to $L_{4-10\ \rm{keV}}<3\times 10^{45}\ \rm{erg\ s^{-1}}$ at $d=377$ Mpc \citep{gcn24218}.

As shown in Figure~\ref{fig:prompt}, the limits on the prompt $\gamma$-ray emission for both events can rule out an on-axis SGRB comparable to the bulk of the energetic cosmological population, which have $E_{\rm\gamma, iso}\approx 10^{51}-10^{52}$ erg. \cite{Song2019} and \cite{Saleem2019} reached a similar conclusion. As indicated in \S\ref{sec:ag}, however, an on-axis orientation is expected in at most a few percent of the cases.

In Figure~\ref{fig:MAXI} we compare the limits from MAXI to the observed X-ray afterglows of cosmological SGRBs, and find that they similarly rule out about half of the observed population, for the fractional areal coverage of each MAXI search.  The same caveat about the rarity of on-axis events applies to these limits as well.  On the other hand, if we compare the MAXI limits to the same off-axis model described in \S\ref{sec:ag}, we find that such a model cannot be constrained. We similarly find that \emph{Swift}/XRT observations carried out for both events \citep{gcn24273,gcn24305,gcn24353} cannot constrain off-axis jets (Figure~\ref{fig:MAXI}).

\section{Conclusions}
\label{sec:conc}

We presented MMT follow-up observations of the first two ALV candidate events in O3 that appear to contain NSs, and are therefore capable of generating EM emission.  Our MMT search targeted 67 and 50 galaxies for S190425z and S190426c, respectively, and did not yield potential counterparts to a limiting magnitude of $i\approx 22.5$.  We further presented our spectroscopic follow-up with SOAR of two candidate optical counterparts from other searches, which revealed unrelated SNe~Ia and IIb.  For comparison we further collated information available from the GCN circulars about other galaxy-targeted and wide-field searches. Due to the large localization areas and volumes of both events all searches were far from complete. Still, a combined total of nearly 100 optical candidates were announced for the two events, and 14 were followed up spectroscopically, revealing normal SNe. 

Parameterizing the efficacy of the searches relative to the brightness of the kilonova associated with GW170817, we find maximal volume coverage of about about $40$\% for S190425z (ZTF) and about $60$\% for S190426c (ZTF plus DECam). Relative to a dimmer kilonova model ($0.01\ M_\odot$ of lanthanide-rich ejecta), a neutron precursor, or a slightly off-axis SGRB, we find that only a few searches (including our MMT observations) reached sufficient depth.  On the other hand, comparing to an on-axis SGRB we find that most searches would have been able to detect such emission, but this is expected in at most a few percent of mergers.

We end with a few general thoughts.  First, the open rapid alerts implemented by the LIGO/Virgo Collaboration in O3 work remarkably well in providing rapid access to sky maps, distance estimates, and rudimentary information about the detections (e.g., FAR). Second, the events considered here indicate that due to duty cycle limitations and the larger detection distances, the localization regions (and volumes) for most events will be much larger than for GW170817, and this is likely to reduce the efficiency of counterpart identification.  Third, despite the larger distances of the GW events at least some searches are capable of reaching the depth necessary to detect GW170817-like kilonovae (if those are common).  Finally, we note that the robustness of the GW detections, as well as the actual properties of the binaries, are difficult to assess with the partial information provided by the LIGO/Virgo Collaboration at the present. In particular, we advocate breaking down the FAR by detector and by search pipeline, which would provide more guidance about the significance of a given event. Furthermore, the chirp mass and the individual component masses potentially provide critical insight about the expected EM signatures \citep{2019arXiv190411995M}. Early release of this additional information is particularly important as the number of detections increases in order to prioritize follow-up and tailor it to the properties of the transient.

\facilities{ADS, MMT (MMTCam), SOAR (Goodman)}
\software{Astropy \citep{astropy}, \texttt{BOXFIT} \citep{vanEerten+11}, Flask, \textit{healpy} \citep{healpy}, JS9 \citep{eric_mandel_2018_1453307}, \texttt{ligo.skymap} \citep{ligo.skymap}, Matplotlib \citep{matplotlib}, MOSFiT \citep{Guillochon+17b}, NumPy \citep{numpy}, Photutils \citep{photutils}, PyGCN \citep{singer_python_2019}, PyZOGY \citep{2017zndo...1043973G}, SciPy \citep{scipy}, SEP \citep{1996A&AS..117..393B,sep}, SNID \citep{Blondin2007}, SQLAlchemy}

\acknowledgements
We thank Dallan Porter and Sean Moran for software development that facilitated our rapid follow-up program at MMT, as well as Ben Kunk for operating the MMT during our observations. We also thank Joey Chatelain for suggesting the use of SkyBoT in our analysis software, and Cristiano Guidorzi for help with the BOXFIT simulations.
The Berger Time-Domain Group is supported in part by NSF grant AST-1714498 and NASA grant NNX15AE50G. G.H.\ thanks the LSSTC Data Science Fellowship Program, which is funded by LSSTC, NSF Cybertraining Grant \#1829740, the Brinson Foundation, and the Moore Foundation; his participation in the program has benefited this work. P.S.C.\ is grateful for support provided by NASA through the NASA Hubble Fellowship grant \#HST-HF2-51404.001-A awarded by the Space Telescope Science Institute, which is operated by the Association of Universities for Research in Astronomy, Inc., for NASA, under contract NAS 5-26555. M.N.\ is supported by a Royal Astronomical Society Research Fellowship. W.F.\ and K.P.\ acknowledge support by the National Science Foundation under award No.\ AST-1814782. I.P.\ acknowledges funding by the Deutsche Forschungsgemeinschaft under grant GE2506/12-1. R.M.\ acknowledges support for this work provided by the National Aeronautics and Space Administration through \textit{Chandra} Award Number DD8-19101A and DDT-18096A issued by the \textit{Chandra} X-ray Center, which is operated by the Smithsonian Astrophysical Observatory for and on behalf of the National Aeronautics Space Administration under contract NAS8-03060. K.D.A.\ is grateful for support provided by NASA through the NASA Hubble Fellowship grant \#HST-HF2-51403.001-A awarded by the Space Telescope Science Institute, which is operated by the Association of Universities for Research in Astronomy, Inc., for NASA, under contract NAS 5-26555. Development of the Boxfit code was supported in part by NASA through grant NNX10AF62G issued through the Astrophysics Theory Program and by the NSF through grant AST-1009863.  This research was supported in part through the computational resources and staff contributions provided for the Quest high-performance computing facility at Northwestern University, which is jointly supported by the Office of the Provost, the Office for Research, and Northwestern University Information Technology. Observations reported here were obtained at the MMT Observatory, a joint facility of the Smithsonian Institution and the University of Arizona, and the Southern Astrophysical Research (SOAR) telescope, which is a joint project of the Minist\'{e}rio da Ci\^{e}ncia, Tecnologia, Inova\c{c}\~{o}es e Comunica\c{c}\~{o}es (MCTIC) do Brasil, the U.S. National Optical Astronomy Observatory (NOAO), the University of North Carolina at Chapel Hill (UNC), and Michigan State University (MSU).

\startlongtable
\begin{deluxetable}{lccccCc}
\tablecaption{Log of MMT Follow-up Observations \label{tab:mmt}}
\tablecolumns{7}
\tablehead{\colhead{Name} & \colhead{R.A.} & \colhead{Decl.} & \colhead{Date} & \colhead{Time} & \colhead{Filter} & \colhead{Limiting Mag.\tablenotemark{a}}}
\startdata
\cutinhead{S190425z}
2MASX J16545364$-$1657072 & 16\textsuperscript{h}54\textsuperscript{m}53\fs65 & $-$16\textsuperscript{d}57\textsuperscript{m}07\fs3 & 2019 Apr 25 & 11:39:23 & g & 22.3 \\
2MASX J16530485$-$1617273 & 16\textsuperscript{h}53\textsuperscript{m}04\fs86 & $-$16\textsuperscript{d}17\textsuperscript{m}27\fs4 & 2019 Apr 25 & 11:40:50 & g & 22.4 \\
2MASX J16571426$-$0613510 & 16\textsuperscript{h}57\textsuperscript{m}14\fs26 & $-$06\textsuperscript{d}13\textsuperscript{m}51\fs0 & 2019 Apr 25 & 11:42:30 & g & 22.4 \\
2MASX J16520774$-$1703135 & 16\textsuperscript{h}52\textsuperscript{m}07\fs75 & $-$17\textsuperscript{d}03\textsuperscript{m}13\fs5 & 2019 Apr 25 & 11:44:22 & g & 22.5 \\
PGC 58929 & 16\textsuperscript{h}45\textsuperscript{m}54\fs64 & $-$23\textsuperscript{d}27\textsuperscript{m}06\fs0 & 2019 Apr 25 & 11:47:33 & g & 22.0 \\
NGC 6234 & 16\textsuperscript{h}51\textsuperscript{m}57\fs34 & +04\textsuperscript{d}23\textsuperscript{m}00\fs8 & 2019 Apr 25 & 11:49:28 & g & 22.4 \\
PGC 59064 & 16\textsuperscript{h}49\textsuperscript{m}33\fs16 & +06\textsuperscript{d}00\textsuperscript{m}58\fs5 & 2019 Apr 25 & 11:51:21 & g & 22.2 \\
PGC 59201 & 16\textsuperscript{h}53\textsuperscript{m}24\fs09 & +04\textsuperscript{d}14\textsuperscript{m}10\fs5 & 2019 Apr 25 & 11:52:39 & g & 22.3 \\
2MASX J16564688$-$0142052 & 16\textsuperscript{h}56\textsuperscript{m}46\fs89 & $-$01\textsuperscript{d}42\textsuperscript{m}05\fs3 & 2019 Apr 25 & 11:54:12 & g & 22.0 \\
2MASX J16504669+0436170 & 16\textsuperscript{h}50\textsuperscript{m}46\fs70 & +04\textsuperscript{d}36\textsuperscript{m}17\fs0 & 2019 Apr 25 & 11:55:46 & g & 22.2 \\
UGC 10426 & 16\textsuperscript{h}30\textsuperscript{m}50\fs16 & +16\textsuperscript{d}15\textsuperscript{m}02\fs5 & 2019 Apr 25 & 11:57:27 & g & 21.9 \\
PGC 90265 & 16\textsuperscript{h}57\textsuperscript{m}26\fs82 & $-$10\textsuperscript{d}11\textsuperscript{m}27\fs9 & 2019 Apr 25 & 11:59:14 & g & 21.7 \\
NGC 6225 & 16\textsuperscript{h}48\textsuperscript{m}21\fs57 & +06\textsuperscript{d}13\textsuperscript{m}22\fs0 & 2019 Apr 25 & 12:00:51 & g & 21.7 \\
2MASX J16462248+0902154 & 16\textsuperscript{h}46\textsuperscript{m}22\fs49 & +09\textsuperscript{d}02\textsuperscript{m}15\fs5 & 2019 Apr 25 & 12:02:05 & g & 21.6 \\
PGC 58705 & 16\textsuperscript{h}39\textsuperscript{m}26\fs39 & +11\textsuperscript{d}12\textsuperscript{m}37\fs7 & 2019 Apr 25 & 12:03:15 & g & 21.8 \\
2MASX J16552449$-$0715255 & 16\textsuperscript{h}55\textsuperscript{m}24\fs50 & $-$07\textsuperscript{d}15\textsuperscript{m}25\fs5 & 2019 Apr 25 & 12:04:57 & g & 21.7 \\
2MASX J16580128$-$0149216 & 16\textsuperscript{h}58\textsuperscript{m}01\fs29 & $-$01\textsuperscript{d}49\textsuperscript{m}21\fs7 & 2019 Apr 25 & 12:06:16 & g & 21.4 \\
2MASX J16590728$-$0544311 & 16\textsuperscript{h}59\textsuperscript{m}07\fs28 & $-$05\textsuperscript{d}44\textsuperscript{m}31\fs2 & 2019 Apr 26 & 08:30:29 & i & 22.2 \\
PGC 58987 & 16\textsuperscript{h}47\textsuperscript{m}24\fs48 & $-$20\textsuperscript{d}08\textsuperscript{m}30\fs3 & 2019 Apr 26 & 08:32:14 & i & 22.2 \\
NGC 6224 & 16\textsuperscript{h}48\textsuperscript{m}18\fs55 & +06\textsuperscript{d}18\textsuperscript{m}43\fs9 & 2019 Apr 26 & 08:34:11 & i & 22.5 \\
NGC 6051 & 16\textsuperscript{h}04\textsuperscript{m}56\fs70 & +23\textsuperscript{d}55\textsuperscript{m}58\fs3 & 2019 Apr 26 & 08:36:37 & i & 21.8 \\
IC 4572 & 15\textsuperscript{h}41\textsuperscript{m}54\fs20 & +28\textsuperscript{d}08\textsuperscript{m}02\fs7 & 2019 Apr 26 & 08:42:22 & i & 22.0 \\
UGC 10320 & 16\textsuperscript{h}18\textsuperscript{m}07\fs32 & +21\textsuperscript{d}03\textsuperscript{m}59\fs0 & 2019 Apr 26 & 08:44:30 & i & 22.5 \\
IC 4569 & 15\textsuperscript{h}40\textsuperscript{m}48\fs35 & +28\textsuperscript{d}17\textsuperscript{m}31\fs4 & 2019 Apr 26 & 08:47:02 & i & 22.1 \\
PGC 57607 & 16\textsuperscript{h}14\textsuperscript{m}57\fs84 & +21\textsuperscript{d}56\textsuperscript{m}17\fs9 & 2019 Apr 26 & 08:50:11 & i & 22.6 \\
PGC 57472 & 16\textsuperscript{h}12\textsuperscript{m}20\fs34 & +23\textsuperscript{d}00\textsuperscript{m}07\fs0 & 2019 Apr 26 & 08:51:41 & i & 22.6 \\
IC 1219 & 16\textsuperscript{h}24\textsuperscript{m}27\fs44 & +19\textsuperscript{d}28\textsuperscript{m}57\fs3 & 2019 Apr 26 & 08:53:28 & i & 22.6 \\
UGC 10412 & 16\textsuperscript{h}29\textsuperscript{m}36\fs14 & +15\textsuperscript{d}39\textsuperscript{m}30\fs4 & 2019 Apr 26 & 08:55:24 & i & 22.6 \\
NGC 6001 & 15\textsuperscript{h}47\textsuperscript{m}45\fs96 & +28\textsuperscript{d}38\textsuperscript{m}30\fs7 & 2019 Apr 26 & 08:57:41 & i & 22.1 \\
PGC 55883 & 15\textsuperscript{h}43\textsuperscript{m}46\fs04 & +28\textsuperscript{d}24\textsuperscript{m}54\fs6 & 2019 Apr 26 & 08:59:39 & i & 22.3 \\
PGC 57645 & 16\textsuperscript{h}15\textsuperscript{m}42\fs15 & +19\textsuperscript{d}38\textsuperscript{m}15\fs0 & 2019 Apr 26 & 09:02:13 & i & 22.6 \\
PGC 59121 & 16\textsuperscript{h}51\textsuperscript{m}21\fs66 & +07\textsuperscript{d}51\textsuperscript{m}44\fs3 & 2019 Apr 26 & 09:04:08 & i & 22.1 \\
PGC 56949 & 16\textsuperscript{h}04\textsuperscript{m}35\fs57 & +25\textsuperscript{d}11\textsuperscript{m}23\fs4 & 2019 Apr 26 & 09:06:30 & i & 22.2 \\
PGC 58860 & 16\textsuperscript{h}44\textsuperscript{m}09\fs28 & +07\textsuperscript{d}26\textsuperscript{m}43\fs0 & 2019 Apr 26 & 09:08:30 & i & 22.0 \\
PGC 57542 & 16\textsuperscript{h}13\textsuperscript{m}46\fs01 & +22\textsuperscript{d}55\textsuperscript{m}08\fs0 & 2019 Apr 26 & 09:10:44 & i & 22.4 \\
UGC 10224 & 16\textsuperscript{h}08\textsuperscript{m}50\fs24 & +22\textsuperscript{d}02\textsuperscript{m}33\fs5 & 2019 Apr 26 & 09:13:36 & i & 22.6 \\
PGC 58097 & 16\textsuperscript{h}25\textsuperscript{m}38\fs08 & +16\textsuperscript{d}27\textsuperscript{m}18\fs0 & 2019 Apr 26 & 09:16:57 & i & 22.6 \\
PGC 1717114 & 16\textsuperscript{h}04\textsuperscript{m}16\fs28 & +24\textsuperscript{d}48\textsuperscript{m}44\fs4 & 2019 Apr 26 & 09:18:57 & i & 22.6 \\
PGC 59239 & 16\textsuperscript{h}54\textsuperscript{m}24\fs03 & $-$09\textsuperscript{d}53\textsuperscript{m}21\fs3 & 2019 Apr 26 & 09:23:22 & i & 22.5 \\
UGC 10260 & 16\textsuperscript{h}11\textsuperscript{m}57\fs88 & +20\textsuperscript{d}55\textsuperscript{m}24\fs5 & 2019 Apr 26 & 09:25:28 & i & 22.7 \\
IC 4570 & 15\textsuperscript{h}41\textsuperscript{m}22\fs56 & +28\textsuperscript{d}13\textsuperscript{m}47\fs3 & 2019 Apr 26 & 09:38:18 & i & 22.7 \\
UGC 10035 & 15\textsuperscript{h}47\textsuperscript{m}36\fs35 & +26\textsuperscript{d}03\textsuperscript{m}49\fs2 & 2019 Apr 26 & 09:44:32 & i & 22.3 \\
PGC 57692 & 16\textsuperscript{h}16\textsuperscript{m}45\fs71 & +19\textsuperscript{d}31\textsuperscript{m}16\fs9 & 2019 Apr 26 & 09:46:51 & i & 22.4 \\
2MASX J16505342$-$1500143 & 16\textsuperscript{h}50\textsuperscript{m}53\fs42 & $-$15\textsuperscript{d}00\textsuperscript{m}14\fs3 & 2019 Apr 26 & 09:49:19 & i & 21.9 \\
NGC 6240 & 16\textsuperscript{h}52\textsuperscript{m}58\fs86 & +02\textsuperscript{d}24\textsuperscript{m}03\fs5 & 2019 Apr 26 & 09:51:27 & i & 22.3 \\
UGC 10360 & 16\textsuperscript{h}23\textsuperscript{m}11\fs34 & +16\textsuperscript{d}55\textsuperscript{m}57\fs4 & 2019 Apr 26 & 09:53:38 & i & 22.6 \\
PGC 55774 & 15\textsuperscript{h}40\textsuperscript{m}36\fs64 & +28\textsuperscript{d}30\textsuperscript{m}44\fs8 & 2019 Apr 26 & 09:56:12 & i & 22.6 \\
PGC 58735 & 16\textsuperscript{h}40\textsuperscript{m}40\fs22 & +14\textsuperscript{d}21\textsuperscript{m}05\fs3 & 2019 Apr 26 & 09:58:50 & i & 22.5 \\
IC 4621 & 16\textsuperscript{h}50\textsuperscript{m}51\fs19 & +08\textsuperscript{d}47\textsuperscript{m}01\fs9 & 2019 Apr 26 & 10:00:50 & i & 22.5 \\
NGC 6075 & 16\textsuperscript{h}11\textsuperscript{m}22\fs57 & +23\textsuperscript{d}57\textsuperscript{m}54\fs5 & 2019 Apr 26 & 10:07:05 & i & 22.5 \\
2MASX J16582619$-$0319463 & 16\textsuperscript{h}58\textsuperscript{m}26\fs20 & $-$03\textsuperscript{d}19\textsuperscript{m}46\fs4 & 2019 Apr 26 & 10:11:42 & i & 22.5 \\
2MASX J16073961+2220315 & 16\textsuperscript{h}07\textsuperscript{m}39\fs62 & +22\textsuperscript{d}20\textsuperscript{m}31\fs5 & 2019 Apr 26 & 10:15:40 & i & 22.5 \\
PGC 58028 & 16\textsuperscript{h}24\textsuperscript{m}15\fs15 & +20\textsuperscript{d}11\textsuperscript{m}01\fs0 & 2019 Apr 26 & 10:17:30 & i & 22.6 \\
PGC 54895 & 15\textsuperscript{h}22\textsuperscript{m}44\fs91 & +29\textsuperscript{d}46\textsuperscript{m}11\fs0 & 2019 Apr 26 & 10:19:52 & i & 22.3 \\
IC 4505 & 14\textsuperscript{h}46\textsuperscript{m}33\fs38 & +33\textsuperscript{d}24\textsuperscript{m}31\fs2 & 2019 Apr 26 & 10:21:52 & i & 22.7 \\
PGC 58768 & 16\textsuperscript{h}41\textsuperscript{m}20\fs90 & +08\textsuperscript{d}54\textsuperscript{m}32\fs6 & 2019 Apr 26 & 10:24:48 & i & 22.5 \\
PGC 55373 & 15\textsuperscript{h}32\textsuperscript{m}46\fs54 & +28\textsuperscript{d}22\textsuperscript{m}01\fs5 & 2019 Apr 26 & 10:27:25 & i & 22.7 \\
IC 4587 & 15\textsuperscript{h}59\textsuperscript{m}51\fs61 & +25\textsuperscript{d}56\textsuperscript{m}26\fs4 & 2019 Apr 26 & 10:29:22 & i & 22.7 \\
UGC 08145 & 13\textsuperscript{h}02\textsuperscript{m}18\fs28 & +32\textsuperscript{d}53\textsuperscript{m}26\fs8 & 2019 Apr 26 & 10:32:45 & i & 22.4 \\
PGC 57293 & 16\textsuperscript{h}09\textsuperscript{m}06\fs46 & +24\textsuperscript{d}52\textsuperscript{m}13\fs1 & 2019 Apr 26 & 10:36:12 & i & 22.5 \\
2MASX J16540875$-$0738073 & 16\textsuperscript{h}54\textsuperscript{m}08\fs76 & $-$07\textsuperscript{d}38\textsuperscript{m}07\fs3 & 2019 Apr 26 & 10:38:59 & i & 22.3 \\
PGC 52138 & 14\textsuperscript{h}35\textsuperscript{m}18\fs42 & +35\textsuperscript{d}07\textsuperscript{m}07\fs7 & 2019 Apr 26 & 10:41:50 & i & 22.2 \\
2MASX J16153554+1927123 & 16\textsuperscript{h}15\textsuperscript{m}35\fs54 & +19\textsuperscript{d}27\textsuperscript{m}12\fs4 & 2019 Apr 26 & 10:44:25 & i & 23.0 \\
PGC 59338 & 16\textsuperscript{h}58\textsuperscript{m}05\fs76 & $-$21\textsuperscript{d}16\textsuperscript{m}26\fs8 & 2019 Apr 26 & 10:46:47 & i & 22.2 \\
UGC 09233 & 14\textsuperscript{h}24\textsuperscript{m}35\fs03 & +35\textsuperscript{d}16\textsuperscript{m}47\fs4 & 2019 Apr 26 & 10:50:02 & i & 23.4 \\
PGC 56421 & 15\textsuperscript{h}56\textsuperscript{m}03\fs87 & +24\textsuperscript{d}26\textsuperscript{m}52\fs7 & 2019 Apr 26 & 10:52:12 & i & 22.6 \\
PGC 1484188 & 16\textsuperscript{h}28\textsuperscript{m}52\fs42 & +15\textsuperscript{d}25\textsuperscript{m}14\fs8 & 2019 Apr 26 & 10:54:31 & i & 22.5 \\
\cutinhead{S190426c}
2MASX J18191810+8807285 & 18\textsuperscript{h}19\textsuperscript{m}18\fs11 & +88\textsuperscript{d}07\textsuperscript{m}28\fs6 & 2019 Apr 27 & 08:38:51 & i & 22.0 \\
2MASX J18215068+8642223 & 18\textsuperscript{h}21\textsuperscript{m}50\fs68 & +86\textsuperscript{d}42\textsuperscript{m}22\fs4 & 2019 Apr 27 & 08:40:44 & i & 22.0 \\
2MASX J18242867+8642139 & 18\textsuperscript{h}24\textsuperscript{m}28\fs67 & +86\textsuperscript{d}42\textsuperscript{m}14\fs0 & 2019 Apr 27 & 08:42:02 & i & 22.1 \\
2MASX J19301513+8540516 & 19\textsuperscript{h}30\textsuperscript{m}15\fs13 & +85\textsuperscript{d}40\textsuperscript{m}51\fs6 & 2019 Apr 27 & 08:44:01 & i & 22.2 \\
2MASX J20412914+8626330 & 20\textsuperscript{h}41\textsuperscript{m}29\fs14 & +86\textsuperscript{d}26\textsuperscript{m}33\fs1 & 2019 Apr 27 & 08:47:16 & i & 22.3 \\
2MASX J20441724+8654219 & 20\textsuperscript{h}44\textsuperscript{m}17\fs24 & +86\textsuperscript{d}54\textsuperscript{m}22\fs0 & 2019 Apr 27 & 08:48:45 & i & 21.4 \\
PGC 3085923 & 20\textsuperscript{h}52\textsuperscript{m}29\fs72 & +86\textsuperscript{d}11\textsuperscript{m}11\fs9 & 2019 Apr 27 & 08:50:07 & i & 22.4 \\
2MASX J20452666+8620428 & 20\textsuperscript{h}45\textsuperscript{m}26\fs66 & +86\textsuperscript{d}20\textsuperscript{m}42\fs9 & 2019 Apr 27 & 08:53:22 & i & 22.2 \\
2MASX J20592695+8454369 & 20\textsuperscript{h}59\textsuperscript{m}26\fs95 & +84\textsuperscript{d}54\textsuperscript{m}37\fs0 & 2019 Apr 27 & 08:55:00 & i & 22.3 \\
2MASX J20110295+4637149 & 20\textsuperscript{h}11\textsuperscript{m}02\fs96 & +46\textsuperscript{d}37\textsuperscript{m}14\fs9 & 2019 Apr 27 & 09:00:56 & i & 22.3 \\
2MASX J20113931+4550035 & 20\textsuperscript{h}11\textsuperscript{m}39\fs32 & +45\textsuperscript{d}50\textsuperscript{m}03\fs6 & 2019 Apr 27 & 09:03:03 & i & 22.3 \\
2MASX J20114858+4657335 & 20\textsuperscript{h}11\textsuperscript{m}48\fs58 & +46\textsuperscript{d}57\textsuperscript{m}33\fs6 & 2019 Apr 27 & 09:04:39 & i & 22.2 \\
2MASX J20132761+4630313 & 20\textsuperscript{h}13\textsuperscript{m}27\fs62 & +46\textsuperscript{d}30\textsuperscript{m}31\fs3 & 2019 Apr 27 & 09:06:17 & i & 22.3 \\
2MASX J20134502+4726333 & 20\textsuperscript{h}13\textsuperscript{m}45\fs03 & +47\textsuperscript{d}26\textsuperscript{m}33\fs4 & 2019 Apr 27 & 09:07:51 & i & 22.3 \\
2MASX J20152058+4555282 & 20\textsuperscript{h}15\textsuperscript{m}20\fs59 & +45\textsuperscript{d}55\textsuperscript{m}28\fs3 & 2019 Apr 27 & 09:10:17 & i & 22.3 \\
2MASX J20201548+4720364 & 20\textsuperscript{h}20\textsuperscript{m}15\fs48 & +47\textsuperscript{d}20\textsuperscript{m}36\fs4 & 2019 Apr 27 & 09:12:40 & i & 22.2 \\
2MASX J20242781+4900526 & 20\textsuperscript{h}24\textsuperscript{m}27\fs81 & +49\textsuperscript{d}00\textsuperscript{m}52\fs7 & 2019 Apr 27 & 09:14:19 & i & 22.1 \\
2MASX J20354336+4953165 & 20\textsuperscript{h}35\textsuperscript{m}43\fs37 & +49\textsuperscript{d}53\textsuperscript{m}16\fs6 & 2019 Apr 27 & 09:15:47 & i & 22.4 \\
2MASX J20224302+5636145 & 20\textsuperscript{h}22\textsuperscript{m}43\fs03 & +56\textsuperscript{d}36\textsuperscript{m}14\fs6 & 2019 Apr 27 & 09:17:53 & i & 22.4 \\
2MASX J20244336+5245430 & 20\textsuperscript{h}24\textsuperscript{m}43\fs37 & +52\textsuperscript{d}45\textsuperscript{m}43\fs0 & 2019 Apr 27 & 09:19:32 & i & 22.3 \\
2MASX J20245359+5610264 & 20\textsuperscript{h}24\textsuperscript{m}53\fs60 & +56\textsuperscript{d}10\textsuperscript{m}26\fs4 & 2019 Apr 27 & 09:20:53 & i & 22.2 \\
2MASX J20260256+5552523 & 20\textsuperscript{h}26\textsuperscript{m}02\fs56 & +55\textsuperscript{d}52\textsuperscript{m}52\fs4 & 2019 Apr 27 & 09:22:19 & i & 22.3 \\
2MASX J20273404+5015483 & 20\textsuperscript{h}27\textsuperscript{m}34\fs04 & +50\textsuperscript{d}15\textsuperscript{m}48\fs3 & 2019 Apr 27 & 09:23:52 & i & 22.3 \\
2MASX J20273859+5353393 & 20\textsuperscript{h}27\textsuperscript{m}38\fs59 & +53\textsuperscript{d}53\textsuperscript{m}39\fs3 & 2019 Apr 27 & 09:25:24 & i & 22.4 \\
2MASX J20281516+5641284 & 20\textsuperscript{h}28\textsuperscript{m}15\fs17 & +56\textsuperscript{d}41\textsuperscript{m}28\fs4 & 2019 Apr 27 & 09:26:51 & i & 22.3 \\
2MASX J20290191+5817016 & 20\textsuperscript{h}29\textsuperscript{m}01\fs92 & +58\textsuperscript{d}17\textsuperscript{m}01\fs6 & 2019 Apr 27 & 09:28:14 & i & 22.3 \\
2MASX J20291160+5219510 & 20\textsuperscript{h}29\textsuperscript{m}11\fs60 & +52\textsuperscript{d}19\textsuperscript{m}51\fs0 & 2019 Apr 27 & 09:29:56 & i & 22.3 \\
2MASX J20300804+5415120 & 20\textsuperscript{h}30\textsuperscript{m}08\fs04 & +54\textsuperscript{d}15\textsuperscript{m}12\fs1 & 2019 Apr 27 & 09:31:48 & i & 22.2 \\
2MASX J20304675+6259395 & 20\textsuperscript{h}30\textsuperscript{m}46\fs76 & +62\textsuperscript{d}59\textsuperscript{m}39\fs5 & 2019 Apr 27 & 09:37:44 & i & 22.4 \\
2MASX J20322187+5812031 & 20\textsuperscript{h}32\textsuperscript{m}21\fs88 & +58\textsuperscript{d}12\textsuperscript{m}03\fs2 & 2019 Apr 27 & 09:39:26 & i & 22.3 \\
2MASX J20334424+5403120 & 20\textsuperscript{h}33\textsuperscript{m}44\fs25 & +54\textsuperscript{d}03\textsuperscript{m}12\fs0 & 2019 Apr 27 & 09:41:06 & i & 22.5 \\
2MASX J20334533+6254178 & 20\textsuperscript{h}33\textsuperscript{m}45\fs34 & +62\textsuperscript{d}54\textsuperscript{m}17\fs9 & 2019 Apr 27 & 09:42:45 & i & 22.4 \\
2MASX J20352447+5759548 & 20\textsuperscript{h}35\textsuperscript{m}24\fs47 & +57\textsuperscript{d}59\textsuperscript{m}54\fs9 & 2019 Apr 27 & 09:44:34 & i & 22.4 \\
2MASX J20354995+6208172 & 20\textsuperscript{h}35\textsuperscript{m}49\fs96 & +62\textsuperscript{d}08\textsuperscript{m}17\fs2 & 2019 Apr 27 & 09:47:17 & i & 22.4 \\
2MASX J20355212+5549587 & 20\textsuperscript{h}35\textsuperscript{m}52\fs13 & +55\textsuperscript{d}49\textsuperscript{m}58\fs8 & 2019 Apr 27 & 09:49:53 & i & 22.3 \\
2MASX J20370886+5756538 & 20\textsuperscript{h}37\textsuperscript{m}08\fs87 & +57\textsuperscript{d}56\textsuperscript{m}53\fs8 & 2019 Apr 27 & 09:51:24 & i & 22.5 \\
2MASX J20373532+5628217 & 20\textsuperscript{h}37\textsuperscript{m}35\fs32 & +56\textsuperscript{d}28\textsuperscript{m}21\fs7 & 2019 Apr 27 & 09:52:51 & i & 22.3 \\
2MASX J20384775+6128473 & 20\textsuperscript{h}38\textsuperscript{m}47\fs75 & +61\textsuperscript{d}28\textsuperscript{m}47\fs4 & 2019 Apr 27 & 09:54:43 & i & 22.3 \\
2MASX J20385946+5351220 & 20\textsuperscript{h}38\textsuperscript{m}59\fs47 & +53\textsuperscript{d}51\textsuperscript{m}22\fs0 & 2019 Apr 27 & 09:56:30 & i & 22.3 \\
2MASX J20391360+6454369 & 20\textsuperscript{h}39\textsuperscript{m}13\fs60 & +64\textsuperscript{d}54\textsuperscript{m}37\fs0 & 2019 Apr 27 & 09:58:12 & i & 22.5 \\
2MASX J20404068+6437003 & 20\textsuperscript{h}40\textsuperscript{m}40\fs69 & +64\textsuperscript{d}37\textsuperscript{m}00\fs3 & 2019 Apr 27 & 09:59:56 & i & 22.2 \\
2MASX J20431546+5424171 & 20\textsuperscript{h}43\textsuperscript{m}15\fs47 & +54\textsuperscript{d}24\textsuperscript{m}17\fs1 & 2019 Apr 27 & 10:02:36 & i & 22.3 \\
2MASX J20450027+6441410 & 20\textsuperscript{h}45\textsuperscript{m}00\fs28 & +64\textsuperscript{d}41\textsuperscript{m}41\fs0 & 2019 Apr 27 & 10:04:16 & i & 22.5 \\
2MASX J20452857+6332249 & 20\textsuperscript{h}45\textsuperscript{m}28\fs58 & +63\textsuperscript{d}32\textsuperscript{m}25\fs0 & 2019 Apr 27 & 10:05:41 & i & 22.3 \\
2MASX J20453196+6157519 & 20\textsuperscript{h}45\textsuperscript{m}31\fs96 & +61\textsuperscript{d}57\textsuperscript{m}52\fs0 & 2019 Apr 27 & 10:07:12 & i & 22.1 \\
2MASX J20482612+6414178 & 20\textsuperscript{h}48\textsuperscript{m}26\fs13 & +64\textsuperscript{d}14\textsuperscript{m}17\fs8 & 2019 Apr 27 & 10:08:31 & i & 22.4 \\
2MASX J20492307+6412331 & 20\textsuperscript{h}49\textsuperscript{m}23\fs08 & +64\textsuperscript{d}12\textsuperscript{m}33\fs2 & 2019 Apr 27 & 10:10:38 & i & 22.4 \\
2MASX J20495907+6207478 & 20\textsuperscript{h}49\textsuperscript{m}59\fs08 & +62\textsuperscript{d}07\textsuperscript{m}47\fs9 & 2019 Apr 27 & 10:12:07 & i & 22.3 \\
2MASX J20515127+6309235 & 20\textsuperscript{h}51\textsuperscript{m}51\fs28 & +63\textsuperscript{d}09\textsuperscript{m}23\fs5 & 2019 Apr 27 & 10:13:43 & i & 22.4 \\
2MASX J20581565+6217178 & 20\textsuperscript{h}58\textsuperscript{m}15\fs66 & +62\textsuperscript{d}17\textsuperscript{m}17\fs8 & 2019 Apr 27 & 10:15:28 & i & 22.3
\enddata
\tablenotetext{a}{\footnotesize These limiting magnitudes correspond to 3 times the sky noise within an aperture of 2.5 times the FWHM, where the sky noise is estimated to be 1.48 times the median absolute deviation of the difference image.}
\end{deluxetable}

\begin{deluxetable*}{lcccccCc}
\tabletypesize{\footnotesize}
\tablecolumns{8}
\tablewidth{0pt}
\tablecaption{Summary of Community Follow-up Observations of S190425z \label{tab:followup190425z}}	
\tablehead{\colhead{GCN}& \colhead{Num. Galaxies} & \colhead{Area (deg$^2$)}& \colhead{Time (UT)} & \colhead{Phase (days)} & \colhead{Limiting Mag.}& \colhead{Filter} & \colhead{Instrument/Group}}
\startdata
GCN24167 \citep{gcn24167} & \nodata & \nodata & 2019 Apr 25 09:14:36 & 0.036 & 17.7 & C & MASTER\tablenotemark{a}\\
GCN24172 \citep{gcn24172} & \nodata & 60 & 2019 Apr 25 09:01:00 & 0.030 & 21 & G & SAGUARO\tablenotemark{b}\\
GCN24175 \citep{gcn24175} & 5 & \nodata & \nodata & \nodata & 22 & B & HET\tablenotemark{c}\\
GCN24179 \citep{gcn24179} & 101 & \nodata & 2019 Apr 25 12:43:52 & 0.034 & 19 & \mathrm{clear} & KAIT\tablenotemark{d}\\
GCN24182 \citep{gcn24182} & 17 & \nodata & 2019 Apr 25 12:06:16 & 0.140 & 21 & g & MMTCam\\
GCN24183 \citep{gcn24183} & 30 & \nodata & 2019 Apr 25 09:38:57 & 0.056 & 20 & i & SQUEAN\tablenotemark{e}\\
GCN24187 \citep{gcn24187} & \nodata & 2401 & 2019 Apr 25 09:12:09 & 0.038 & 16.75 & J & Gattini-IR\\
GCN24188 \citep{gcn24188} & 13 & \nodata & 2019 Apr 25 10:17:06 & 0.083 & 19 & R & LOAO\tablenotemark{f}\\
GCN24190 \citep{gcn24190} & \nodata & 79 & 2019 Apr 25 12:40:09 & 0.182 & 18 & \nodata & Xinglong-Schmidt\\
GCN24191 \citep{gcn24191} & \nodata & 4327 & 2019 Apr 25 09:19:07 & 0.042 & 20.4 & g,r & ZTF\tablenotemark{g}\\
GCN24192 \citep{gcn24192} & 154 & \nodata & 2019 Apr 25 11:46:00 & 0.144 & 23.5 & r & FOCAS\tablenotemark{h}\\
GCN24193 \citep{gcn24193} & 27 & \nodata & 2019 Apr 25 12:27:23 & 0.173 & 20 & R & LOT\tablenotemark{i}\\
GCN24197 \citep{gcn24197} & \nodata & 2652 & 2019 Apr 25 09:18:02 & 0.042 & 19.5 & o & ATLAS\tablenotemark{j}\\
GCN24198 \citep{gcn24198} & 10 & \nodata & 2019 Apr 25 10:12:00 & 0.079 & 20.8 & r & KPED\tablenotemark{k}\\
GCN24207 \citep{gcn24207} & 21 & \nodata & 2019 Apr 25 13:42:17 & 0.191 & 21.7 & g,i,r & LasCumbres-SSO\tablenotemark{l}\\
GCN24210 \citep{gcn24210} & \nodata & 1258 & 2019 Apr 25 09:39:48 & 0.057 & 21.7 & i & Pan-STARRS\tablenotemark{m}\\
GCN24216 \citep{gcn24216} & 120 & \nodata & 2019 Apr 25 12:28:00 & 0.174 & \nodata & R & KMTNet\tablenotemark{n}\\
GCN24224 \citep{gcn24224} & \nodata & 2134 & 2019 Apr 25 20:38:00 & 0.514 & 20.1 & L & GOTO\tablenotemark{o}\\
GCN24225 \citep{gcn24225} & 19 & \nodata & 2019 Apr 25 23:15:41 & 0.471 & 21.4 & g,i,r & LasCumbres-SAAO\tablenotemark{p}\\
GCN24227 \citep{gcn24227} & \nodata & 123 & 2019 Apr 25 19:43:51 & 0.277 & 17 & \nodata & TAROT-GRANDMA\tablenotemark{q}\\
GCN24238 \citep{gcn24238} & 23 & \nodata & 2019 Apr 26 11:40:00 & 0.808 & \nodata & \nodata & RATIR\tablenotemark{r}\\
GCN24239 \citep{gcn24239} & 128 & \nodata & \nodata & \nodata & \nodata & w & COATLI\tablenotemark{s}\\
GCN24244 \citep{gcn24244} & 50 & \nodata & 2019 Apr 26 10:54:31 & 1.009 & 21 & i & MMTCam\\
GCN24256 \citep{gcn24256} & 119 & \nodata & 2019 Apr 26 01:03:45 & 0.292 & 19.2 & \nodata & GRANDMA\tablenotemark{t}\\
GCN24270 \citep{gcn24270} & 63 & \nodata & 2019 Apr 25 09:57:38 & 0.069 & 20.5 & \mathrm{clear} & BOOTES-5/JGT\tablenotemark{u}\\
GCN24274 \citep{gcn24274} & 58 & \nodata & \nodata & \nodata & 20 & R & LOT\tablenotemark{i}\\
GCN24285 \citep{gcn24285} & \nodata & 675 & \nodata & \nodata & 20.983 & V,R & CNEOST\tablenotemark{v}\\
GCN24309 \citep{gcn24309} & \nodata & 5000 & 2019 Apr 25 09:18:02 & 0.042 & 18.25 & g & ASAS-SN\tablenotemark{w}\\
GCN24311 \citep{gcn24311} & \nodata & 4950 & \nodata & \nodata & 21 & g,r & ZTF\tablenotemark{g}\\
GCN24315 \citep{gcn24315} & 80 & \nodata & 2019 Apr 26 13:58:39 & 0.164 & 16.86 & R & GWAC-F60A\tablenotemark{x}\\
GCN24353 \citep{gcn24353} & 408 & \nodata & 2019 Apr 25 23:45:27 & 0.644 & 21.1 & u & UVOT\tablenotemark{y}\\
GCN24367 \citep{gcn24367} & 5 & \nodata & 2019 Apr 25 21:36:00 & 0.554 & 21.5 & r & HET\tablenotemark{c}\\
\enddata
\tablecomments{Compilation of all public follow-up searches reported for S190425z.  The list includes both galaxy-targeted and wide-field searches, with their respective number of galaxies observed or area covered in square degrees. The start time of each observation, the approximate limiting magnitude of each instrument, and the filters used by each survey are also shown. The entries are sorted by GCN Circular number.\vspace{-6pt}}
\tablenotetext{a}{\vspace{-6pt}Mobile Astronomical System of Telescope Robots}
\tablenotetext{b}{\vspace{-6pt}Searches After Gravitational Waves Using Arizona's Observatories}
\tablenotetext{c}{\vspace{-6pt}Hobby--Eberly Telescope}
\tablenotetext{d}{\vspace{-6pt}Katzmann Automatic Imaging Telescope}
\tablenotetext{e}{\vspace{-6pt}Spectral Energy Distribution (SED) Camera for Quasars in Early Universe}
\tablenotetext{f}{\vspace{-6pt}Lemonsan Optical Astronomical Observatory}
\tablenotetext{g}{\vspace{-6pt}Zwicky Transient Facility}
\tablenotetext{h}{\vspace{-6pt}Faint Object Camera And Spectrograph}
\tablenotetext{i}{\vspace{-6pt}Lulin One-meter Telescope}
\tablenotetext{j}{\vspace{-6pt}Asteroid Terrestrial-impact Last Alert System}
\tablenotetext{k}{\vspace{-6pt}Kitt Peak Electron-Multiplying Charge-Coupled Device (EMCCD) Demonstrator}
\tablenotetext{l}{\vspace{-6pt}Las Cumbres Observatory node at Siding Spring Observatory}
\tablenotetext{m}{\vspace{-6pt}Panoramic Survey Telescope and Rapid Response System}
\tablenotetext{n}{\vspace{-6pt}Korean Microlensing Telescope Network}
\tablenotetext{o}{\vspace{-6pt}Gravitational-wave Optical Transient Observer}
\tablenotetext{p}{\vspace{-6pt}Las Cumbres Observatory node at the South African Astronomical Observatory}
\tablenotetext{q}{\vspace{-6pt}T\'elescopes \`a Action Rapide pour les Objets Transitoires, part of GRANDMA\tablenotemark{t}}
\tablenotetext{r}{\vspace{-6pt}Reionization and Transients IR Camera}
\tablenotetext{s}{\vspace{-6pt}Corrector de \'Optica \'Activa y de Tilts al L\'imite de Difracci\'on}
\tablenotetext{t}{\vspace{-6pt}Global Rapid Advanced Network Devoted to the Multi-messenger Addicts}
\tablenotetext{u}{\vspace{-6pt}Burst Observer and Optical Transient Exploring System 5, also known as the Javier Gorosabel Telescope}
\tablenotetext{v}{\vspace{-6pt}China Near Earth Object Survey Telescope}
\tablenotetext{w}{\vspace{-6pt}All-Sky Automated Survey for Supernovae}
\tablenotetext{x}{\vspace{-6pt}60 cm telescope at Xinglong Observatory}
\tablenotetext{y}{\vspace{-6pt}UV/Optical Telescope}
\end{deluxetable*}

\begin{deluxetable*}{lcccccCc}
\tabletypesize{\footnotesize}
\tablecolumns{8}
\tablewidth{0pt}
\tablecaption{Summary of Community Follow-up Observations of S190426c\label{tab:followup190426c}}	
\tablehead{\colhead{GCN}& \colhead{Num. Galaxies} & \colhead{Area (deg$^2$)}& \colhead{Time (UT)} & \colhead{Phase (days)} & \colhead{Limiting Mag.}& \colhead{Filter} & \colhead{Instrument/Group}}
\startdata
GCN24236 \citep{gcn24236} & \nodata & \nodata & 2019 Apr 26 16:25:18 & 0.037 & 18.3 & \mathrm{Clear} & MASTER\tablenotemark{a}\\
GCN24247 \citep{gcn24247} & 19 & \nodata & 2019 Apr 26 16:27:00 & 0.045 & \nodata & \nodata & SNU\tablenotemark{b}\\
GCN24257 \citep{gcn24257} & \nodata & 830 & 2019 Apr 26 23:42:21 & 0.318 & 22.9 & r,z & DECam\tablenotemark{c}\\
GCN24258 \citep{gcn24258} & \nodata & 7.5 & \nodata & \nodata & 20.5 & r & GIT\tablenotemark{d}\\
GCN24278 \citep{gcn24278} & 5 & \nodata & \nodata & \nodata & 22 & B & HET\tablenotemark{e}\\
GCN24281 \citep{gcn24281} & 10 & \nodata & 2019 Apr 26 17:19:27 & 0.082 & 18.5 & \mathrm{Clear} & NEXT-0.6m\tablenotemark{f}\\
GCN24283 \citep{gcn24283} & \nodata & 4340 & 2019 Apr 27 05:45:00 & 0.599 & 22 & g,r & ZTF\tablenotemark{g}\\
GCN24284 \citep{gcn24284} & \nodata & 2200 & 2019 Apr 27 03:31:00 & 0.506 & 15 & J & Gattini-IR\\
GCN24286 \citep{gcn24286} & \nodata & 774 & 2019 Apr 26 16:38:56 & 0.053 & 20.667 & \nodata & CNEOST\tablenotemark{h}\\
GCN24289 \citep{gcn24289} & 247 & \nodata & 2019 Apr 27 06:26:42 & 0.525 & 19 & \nodata & KAIT\tablenotemark{i}\\
GCN24291 \citep{gcn24291} & \nodata & 755 & 2019 Apr 26 20:38:00 & 0.220 & 19.9 & L & GOTO\tablenotemark{j}\\
GCN24292 \citep{gcn24292} & 50 & \nodata & 2019 Apr 27 10:15:28 & 0.720 & 21 & i & MMTCam\\
GCN24298 \citep{gcn24298} & 98 & \nodata & 2019 Apr 27 11:46:00 & 0.511 & 19.5 & w & COATLI\tablenotemark{k}\\
GCN24299 \citep{gcn24299} & 39 & \nodata & \nodata & \nodata & 15.4,17.6,---,17.7,20.0 & H,J,K,R,\mathrm{clear} & J-GEM\tablenotemark{l}\\
GCN24300 \citep{gcn24300} & 22 & \nodata & 2019 Apr 27 11:41:00 & 0.506 & 21.2 & i,g,Y,H & RATIR\tablenotemark{m}\\
GCN24310 \citep{gcn24310} & \nodata & 384 & 2019 Apr 27 11:01:00 & 0.612 & \nodata & w & DDOTI/OAN\tablenotemark{n}\\
GCN24316 \citep{gcn24316} & \nodata & 12.3 & \nodata & \nodata & 20.6 & \nodata & GIT\tablenotemark{d}\\
GCN24322 \citep{gcn24322} & 23 & \nodata & 2019 Apr 27 09:14:26 & 0.745 & 20.8 & R & LOAO\tablenotemark{o}\\
GCN24323 \citep{gcn24323} & \nodata & 973 & 2019 Apr 26 16:22:17 & 0.042 & 18.25 & g & ASAS-SN\tablenotemark{p}\\
GCN24327 \citep{gcn24327} & \nodata & 43.1 & 2019 Apr 27 02:32:20 & 0.263 & 19.6 & r & GRANDMA\tablenotemark{q}\\
GCN24329 \citep{gcn24329} & \nodata & 1900 & 2019 Apr 28 03:28:00 & 1.504 & 15 & J & Gattini-IR\\
GCN24331 \citep{gcn24331} & \nodata & 4420 & \nodata & \nodata & 22 & g,r & ZTF\tablenotemark{g}\\
GCN24336 \citep{gcn24336} & 17 & \nodata & 2019 Apr 28 09:59:34 & 1.776 & 19.6 & R & LOAO\tablenotemark{o}\\
GCN24340 \citep{gcn24340} & \nodata & 5.0 & 2019 Apr 26 22:44:49 & 0.295 & 19.4 & r & Asiago-Schmidt\\
GCN24346 \citep{gcn24346} & 48 & \nodata & 2019 Apr 27 14:11:17 & 0.949 & 18.40 & R_\mathrm{C} & Yaoan\\
GCN24353 \citep{gcn24353} & 959 & \nodata & 2019 Apr 26 17:44:46 & 0.099 & 21.1 & u & UVOT\tablenotemark{r}\\
\enddata
\tablecomments{See notes in Table~\ref{tab:followup190425z}.\vspace{-6pt}}
\tablenotetext{a}{\vspace{-6pt}Mobile Astronomical System of Telescope Robots}
\tablenotetext{b}{\vspace{-6pt}Seoul National University telescope}
\tablenotetext{c}{\vspace{-6pt}Dark Energy Camera}
\tablenotetext{d}{\vspace{-6pt}Global Relay of Observatories Watching Transients Happen (GROWTH) India Telescope}
\tablenotetext{e}{\vspace{-6pt}Hobby--Eberly Telescope}
\tablenotetext{f}{\vspace{-6pt}Ningbo Bureau of Education and Xinjiang Observatory 0.6 m telescope}
\tablenotetext{g}{\vspace{-6pt}Zwicky Transient Facility}
\tablenotetext{h}{\vspace{-6pt}China Near Earth Object Survey Telescope}
\tablenotetext{i}{\vspace{-6pt}Katzmann Automatic Imaging Telescope}
\tablenotetext{j}{\vspace{-6pt}Gravitational-wave Optical Transient Observer}
\tablenotetext{k}{\vspace{-6pt}Corrector de \'Optica \'Activa y de Tilts al L\'imite de Difracci\'on}
\tablenotetext{l}{\vspace{-6pt}Japanese Collaboration for Gravitational-wave EM Follow-up}
\tablenotetext{m}{\vspace{-6pt}Reionization and Transients IR Camera}
\tablenotetext{n}{\vspace{-6pt}Deca-Degree Optical Transient Imager at the Observatorio Astron\'omico Nacional (M\'exico)}
\tablenotetext{o}{\vspace{-6pt}Lemonsan Optical Astronomical Observatory}
\tablenotetext{p}{\vspace{-6pt}All-Sky Automated Survey for Supernovae}
\tablenotetext{q}{\vspace{-6pt}Global Rapid Advanced Network Devoted to the Multi-messenger Addicts}
\tablenotetext{r}{\vspace{-6pt}UV/Optical Telescope}
\end{deluxetable*}

\end{document}